\documentclass[apj]{emulateapj}
\slugcomment{{Accepted for publication in the Astrophysical Journal}}

\begin{document}

\title{\textit{Chandra} and \textit{XMM-Newton} study of the supernova remnant Kes~73 hosting the magnetar 1E 1841--045}
\author{Harsha S. Kumar\altaffilmark{1}, Samar Safi-Harb\altaffilmark{1,2}, Patrick O. Slane\altaffilmark{3} \& E. V. Gotthelf\altaffilmark{4}}
\altaffiltext{1}{Department of Physics \& Astronomy, University of Manitoba, Winnipeg, MB R3T 2N2, Canada; harsha@physics.umanitoba.ca, samar@physics.umanitoba.ca}
\altaffiltext{2}{Canada Research Chair}
\altaffiltext{3}{Harvard-Smithsonian center for Astrophysics, Cambridge, MA 02138, USA; slane@cfa.harvard.edu}
\altaffiltext{4}{Columbia Astrophysics Laboratory, Columbia university, New York, NY 10027, USA; eric@astro.columbia.edu}

\begin{abstract}

We present a \textit{Chandra} and \textit{XMM-Newton} study of the supernova remnant (SNR) Kes 73 hosting the anomalous X-ray pulsar 1E~1841$-$045. The Chandra image reveals clumpy structures across the remnant with enhanced emission along the western rim. The X-ray emission fills the radio shell and spatially correlates with the infrared image. The global X-ray spectrum is described by a two-component thermal model with a column density $N_{\rm H}$ = 2.6$^{+0.4}_{-0.3}$$\times$10$^{22}$~cm$^{-2}$ and a total luminosity of $L_X$ = 3.3$^{+0.7}_{-0.5}$$\times$10$^{37}$~ergs~s$^{-1}$ (0.5--10 keV, at an assumed distance of 8.5 kpc).  The soft component is characterized by a temperature $kT_s$ = 0.5$^{+0.1}_{-0.2}$~keV, a high ionization timescale, and enhanced Si and S abundances suggesting emission that is dominated by shocked ejecta. The hard component has a temperature $kT_h$ = 1.6$^{+0.8}_{-0.7}$~keV, a relatively low ionization timescale, and mostly solar abundances suggesting emission that is dominated by interstellar/circumstellar shocked material. A spatially resolved spectroscopy study reveals no significant variations in the spectral properties. We infer an SNR age ranging between 750 yr and 2100 yr, an explosion energy of 3.0$^{+2.8}_{-1.8}$$\times$10$^{50}$~ergs and a shock velocity of (1.2 $\pm$ 0.3)$\times$10$^{3}$~km~s$^{-1}$ (under the Sedov phase assumption). We also discuss the possible scenario for Kes~73 expanding into the late red supergiant wind phase of its massive progenitor. Comparing the inferred metal abundances to core-collapse nucleosynthesis model yields, we estimate a progenitor mass $\gtrsim$20 $M_{\sun}$, adding a candidate to the growing list of highly magnetized neutron stars proposed to be associated with very massive progenitors.

\end{abstract}


\section{Introduction}
\label{1}

The Galactic supernova remnant (SNR) Kes~73 (G27.4+0.0) has been classified as a shell-type SNR and hosts the anomalous X-ray pulsar (AXP) 1E 1841$-$045 (Helfand et al. 1994; Vasisht \& Gotthelf 1997). Radio studies of the remnant show an incomplete, small diameter ($\sim$5$^{\prime}$), shell characterized by a steep spectral index ($\alpha$ $\sim$ 0.68), a flux density of 6~Jy at 1~GHz, and with no detection of the central compact source with a flux limit $F_{6 cm}$ $<$ 0.45 mJy and $F_{20 cm}$ $<$ 0.60 mJy (Green 2009; Kriss et al. 1985; Helfand et al. 1994). The infrared studies, carried out as part of the Galactic SNR surveys, clearly detected the remnant in the 24 $\mu$m band (Wachter et al. 2007; Carey et al. 2009; Pinheiro Goncalves et al. 2011). The infrared emission originating from the SNR with an estimated dust mass of 0.11~$M_{\sun}$ has been explained by Pinheiro Goncalves et al. (2011) as likely due to grains heated by collisions in the hot plasma, while Carey et al. (2009) suggested that the 24 $\mu$m emission may be mostly due to nebular emission lines such as [O IV] (25.89 $\mu$m) and [Fe II] (25.99 $\mu$m) with a small contribution to the continuum emission from dust produced in the remnant.

The SNR Kes~73/AXP system was previously observed and studied using many X-ray observatories including \textit{Einstein}, \textit{ROSAT}, \textit{ASCA}, \textit{Chandra}, and \textit{XMM-Newton}. The \textit{ASCA} spectrum was described by a thermal bremsstrahlung model with $kT$ $\sim$ 0.6~keV plus Gaussian emission lines with evidence for enhanced Mg, and possibly O and Ne, abundances, with the SNR age estimated to be $\leq$2000~years (Gotthelf \& Vasisht 1997).  A \textit{Chandra} observation of the SNR was briefly studied by Morii et al. (2010) in connection with the \textit{Suzaku} studies of AXP~1E~1841$-$045, where the entire SNR spectrum was modeled by a VSEDOV model. The \textit{XMM-Newton} study of the remnant was performed by Vink \& Kuiper (2006) using the MOS data.  They fitted the spectrum using either a two-component non-equilibrium ionization (NEI) model under the SPEX fitting package (Kaastra \& Mewe 2000) or a one-component VSEDOV model, both of which  yielded solar metal abundances.  Their study, primarily targeted to determine the supernova explosion energy, argued against the millisecond proto-neutron star model for magnetar formation. Recently, Lopez et al. (2011) also reported X-ray results on  Kes~73 as part of a survey study aimed at typing SNRs using their X-ray morphology as well as to set observational constraints on the hydrodynamical models. Their study, which made use of one of the available \textit{Chandra} observations of Kes~73, yielded a mean temperature of $kT$ = 0.84 $\pm$ 0.49~keV using a one-component VPSHOCK model and detected enhanced abundances from Mg, Si, and S for some of the small scale regions extracted from within the remnant. 

The above mentioned X-ray studies lack a detailed imaging and spectroscopic analysis of the remnant using all available data. Hence, in the following work, we extend the earlier studies to include multi-wavelength data and in particular, the previously unpublished \textit{Chandra} and \textit{XMM-Newton} observations. We provide a detailed X-ray imaging-spectroscopic analysis to investigate the SNR's multi-wavelength morphology (radio, infrared, and X-rays), map the spectral parameters across the remnant, revisit the supernova explosion properties, and address the mass of its progenitor star. 

The latter goal is motivated by addressing the nature of the progenitors and the environment of highly magnetized neutron stars (with a surface dipole magnetic field $B$ $\gtrsim$ 4$\times$10$^{13}$~G) through studying their hosting SNRs (see Safi-Harb \& Kumar 2012). Kes~73 is one of only a few SNRs associated with high-$B$ pulsars. The X-ray emission from its associated AXP,  the slowest  ($P$ = 11.8 s) known magnetar with a characteristic age of $\sim$4.7~kyr, was first decoupled from the diffuse emission from the SNR using \textit{ASCA} (Helfand et al. 1994; Gotthelf \& Vasisht 1997). This source has been manifesting itself as a quiescent magnetar since the time of its discovery until recently, when it exhibited the first magnetar-like burst caught by the \textit{Swift} X-ray observatory (Beardmore et al. 2010; Kumar \& Safi-Harb 2010) followed by a few more bursts  (Lin et al. 2011), further establishing its magnetar nature with $B$ $\sim$ 7$\times$10$^{14}$~G.  Although the AXP's emission will not be further discussed in this paper, the study of SNR~Kes~73, because of its relative youth and brightness, provides a unique opportunity to shed light on the progenitor of highly magnetized neutron stars.

The paper is organized as follows: Section 2 describes the \textit{Chandra} and \textit{XMM-Newton} observations and data reduction. In Sections 3 and 4, we present the details of the X-ray imaging analysis and spectral fitting of the remnant, respectively. Section 5 discusses the results and finally, we summarize our study in Section 6.

\section{Observations \& Data Reduction}
\label{2}

\subsection{\textit{Chandra}}
\label{2.1}

The SNR Kes~73 was observed with the Advanced CCD Imaging Spectrometer (ACIS) onboard the \textit{Chandra X-ray Observatory} on 2000 July 23 (ObsID: 729) and 2006 July 30 (ObsID: 6732). The SNR coordinates were positioned at the aimpoint of the back-illuminated ACIS-S3 chip. Data were taken in full-frame timed-exposure mode and the CCD temperature was $-$120$^{\circ}$ C with a CCD frame readout time of 3.2 s. The data were reduced using the standard \textit{Chandra} Interactive Analysis of Observations (CIAO) version 4.3 routines\footnote{http://cxc.harvard.edu/ciao}. We reprocessed the event files (from level 1 to level 2) to remove pixel randomization and to correct for CCD charge transfer efficiencies.  The bad grades were filtered out and good time intervals were reserved. The resulting total effective exposure for the two observations after data processing is 54.1~ks (see Table~1).
 
\subsection{XMM-Newton}
\label{2.2}

\textit{XMM-Newton} observed Kes~73 on 2002 October 5 and 7 (ObsIDs: 0013340101 and 0013340201) with the European Photon Imaging Camera (EPIC), which has one pn (Str\"{u}der et al. 2001) and two MOS cameras (Turner et al. 2001) covering the energy range between 0.2 and 12 keV with a energy resolution of 0.15 keV at 1 keV. Their on-axis angular resolution is around 6$\arcsec$ FWHM and 15$\arcsec$ half-power diameter with a field of view of 30$\arcmin$. The EPIC-pn and MOS cameras were operated in Full Frame mode with a time resolution of 73 ms and 2.6~s, respectively. We analyzed the data from both the MOS and pn instruments using the $\textit{XMM}$ Science Analysis System (SAS)\footnote{http://xmm.esa.int/sas/} version 10.0.0 and the most recent calibration files. The event files were created from observational data files (ODFs) using the SAS tasks \textit{epchain} and \textit{emchain}. The events were then filtered to retain only patterns 0 to 4 for the pn data (0.2--15 keV energy band) and patterns 0 to 12 for the MOS data (0.2--12 keV energy range). The data were screened to remove spurious events and time intervals with heavy proton flaring by inspecting the light curves for each instrument separately at energies above 10 keV. The light curves were created with 100~s bins and the bins with count-rates greater than 0.35 and 0.4 counts s$^{-1}$ were rejected for MOS1/2 and pn, respectively. The resulting total effective exposure for the MOS1+2 and pn cameras for the two observations is $\sim$20.3~ks and $\sim$6.7~ks, respectively (see Table~1).

\begin{table*}[ht]
\caption{X-ray observation log of SNR Kes~73}
\begin{tabular}{l l l l l c}
\hline\hline Satellite & ObsID & Date of  &  Detector & Total & Effective  \\
& & observation & & Exposure (ks) & Exposure (ks)\\
\hline\hline
\textit{Chandra} & 729 & 23 Jul 2000 & ACIS-S3 & 29.6 &   29.3 \\
\textit{XMM-Newton} &  0013340101 & 5 Oct 2002 & MOS1 & 5.78 & 3.75\\
& & & MOS2 & 5.77 & 3.93 \\
&  & & pn & 3.86 & 2.34 \\
\textit{XMM-Newton} & 0013340201 & 7 Oct 2002 & MOS1 & 6.37 & 6.30 \\
& & & MOS2 & 6.37 & 6.31\\
& & & pn & 4.43 & 4.36 \\
\textit{Chandra} & 6732 & 30 Jul 2006 & ACIS-S3 & 25.2 &  24.8 \\
\hline
\end{tabular}
\end{table*}

\section{X-ray imaging analysis}
\label{3}

In this section, we describe the imaging analysis of Kes~73, observed with both the \textit{Chandra} and \textit{XMM-Newton} satellites. Its associated magnetar 1E~1841$-$045 has a position centered at $\alpha_{J2000}$ = 18$^{h}$41$^{m}$19$^{s}.343$ and $\delta_{J2000}$ = $-$04$^{\circ}$56\hbox{$^{\prime}$}11\hbox{$^{\prime\prime}$}.16 (J2000) with a 1$\sigma$ error of 0$\arcsec$.3 (Wachter et al. 2004).  Due to \textit{Chandra}'s superior angular resolution compared to that of \textit{XMM-Newton}, we concentrate our arcsecond-scale narrow-band and broadband imaging analysis on the ACIS data. The \textit{XMM-Newton} data are additionally used for a combined spectral analysis on arcminute-scale.

We first divided the two \textit{Chandra} observations into individual images in the soft (0.5--1.7 keV; red), medium (1.7--2.7 keV; green), and hard (2.7--7.0 keV; blue) energy bands (see Figure~1). The choice of the energy bands was mainly based on our spectral analysis (Section~4), such that they cover the line emission regions for Mg (1.27--1.42 keV), Si (1.72--1.95 keV), and S (2.31--2.61 keV). The 7~keV upper bound for the hard-band image is due to the poor signal to noise ratio above this energy. The resulting images were subsequently exposure-corrected and the two \textit{Chandra} observations were merged.  Each of the three energy images was smoothed using a Gaussian function with $\sigma$~=~2$\arcsec$, and then the three images were combined to produce the tri-color (RGB) image shown in Figure 1 (bottom).  Images generated from \textit{XMM-Newton} data (not shown) are consistent in morphology to those of the \textit{Chandra} images, but at a lower spatial resolution.

In X-rays, the remnant has a nearly circular morphology with an apparent diameter of $\sim$4.5$\arcmin$.  The \textit{Chandra} images (Figure~1) reveal several small scale clumpy and bright knotty structures filling the SNR interior that are more noticeable in the soft energy band where line emission dominates.  An incomplete outer shell-like feature runs from east to south. The interior region to the east of the AXP shows bright diffuse emission and forms an interesting $W$-shaped structure, which extends towards the southern side of the SNR to merge with the bright western limb. This $W$-shaped arc is clearly seen in all the three energy band images, thus appearing white in the RGB images. The X-ray images further show a faint filamentary structure inside the remnant's bright western limb, to the west of the AXP. The northern portion of the SNR is relatively faint in X-rays, with no distinct shell-like feature.  In Section 5.1, we further discuss the morphology and environment of Kes~73 in comparison with the radio and infrared images.

\begin{figure*} [t]
%\vspace{-0.8cm}
\center
\includegraphics[width=0.8\textwidth]{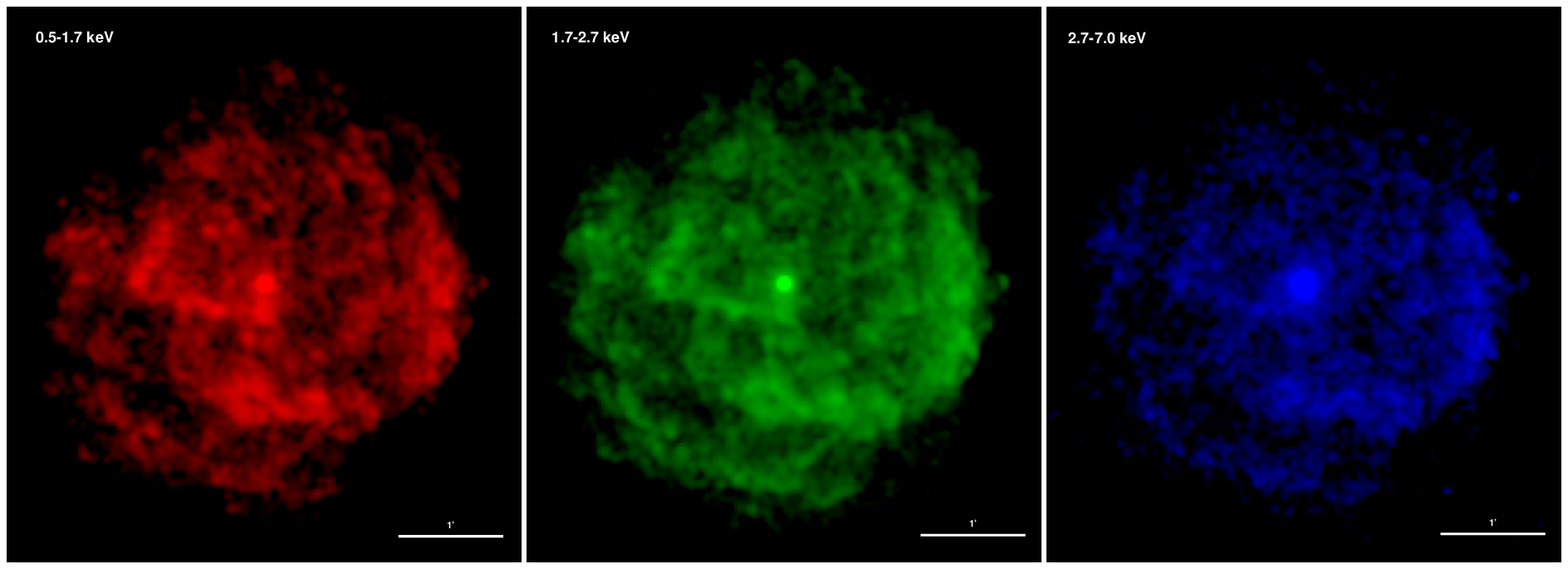}
\includegraphics[width=0.9\textwidth]{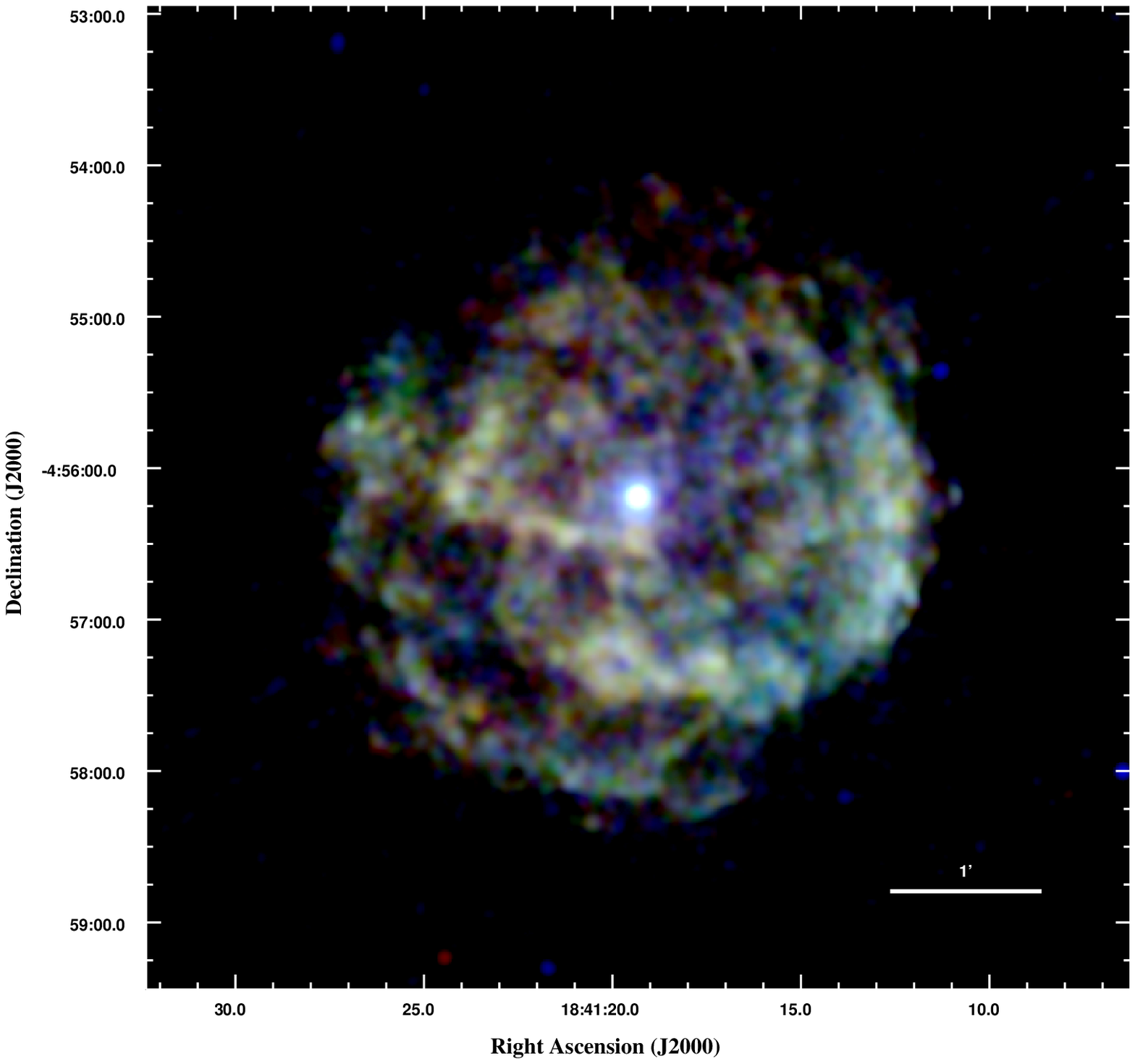}
\caption{\textit{Top}: X-ray images of the SNR Kes~73  in different energy bands: 0.5--1.7 keV (soft; left) in red, 1.7--2.7 keV (medium; center) in green and 2.7--7.0 keV (hard; right) in blue. \textit{Bottom}: \textit{Chandra} ACIS-S3 tri-color image created by combining the soft, medium, and hard energy band images. The central bright source is the AXP 1E~1841$-$045. All the images (in logarithmic scale), created by merging the two \textit{Chandra} observations, are exposure-corrected and smoothed using a $\sigma$ = 2$\arcsec$ Gaussian. North is up and East is to the left.}
\end{figure*}

\begin{figure*}[ht]
\includegraphics[width=\textwidth]{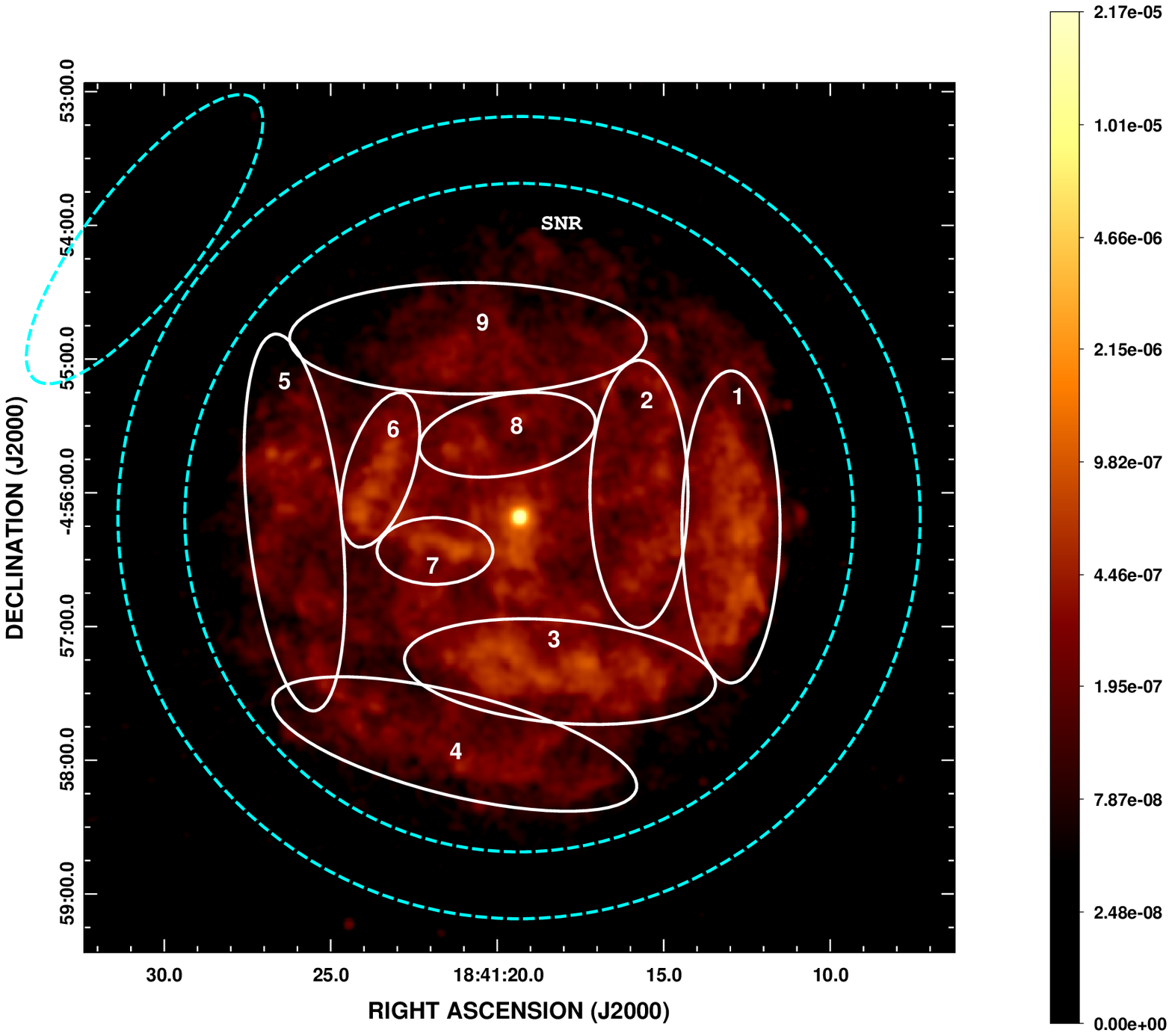}
\caption{\textit{Chandra} ACIS-S3 broadband (0.5--7.0 keV) exposure-corrected intensity image of SNR Kes~73, smoothed using a Gaussian with $\sigma$ = 5$\arcsec$. The color bar is in logarithmic scale and the brightness in units of photon~cm$^{-2}$~s$^{-1}$~arcsec$^{-2}$. The SNR diffuse emission regions selected for a spatially resolved spectroscopic study are overlaid in white for regions~1--9.  The global SNR encloses the region inside the inner annulus (150$\arcsec$) with the contribution from the pulsar (central bright source in the images) and other point sources removed.  The background (in cyan) was extracted from nearby source-free regions within the \textit{Chandra} chip, as shown by the dashed ellipse in the northeast (for regions~1--9) and the annulus (for the global SNR spectrum). The same source and background regions were extracted from the \textit{XMM-Newton} data). The image contrast has been reduced to display the regions and numbers. North is up and East is to the left.}
\end{figure*}

\section{Spectral analysis}
\label{4}

In the following, we perform a spatially resolved spectroscopic analysis of the SNR combining all the \textit{Chandra} and the \textit{XMM-Newton} data. For spectra extraction, we used the XMMSAS task \textit{evselect} for the pn and MOS1/2 cameras, and the CIAO specific command \textit{specextract} for the ACIS-S3 data. Ancillary response files (ARFs) and redistribution matrix files (RMFs) were calculated for the corresponding detector regions. The contributions from point sources within the detector chips were removed prior to the extraction of the spectra. The spectra were extracted in the 0.5--10.0 keV energy range for the \textit{XMM-Newton} data, whereas, we ignored the energy range above 7.0 keV for the \textit{Chandra} observations due to the poor signal-to-noise ratio.    

 \begin{table}[tb] 
\caption{Regions of diffuse emission extracted from the SNR Kes~73 using \textit{Chandra} and \textit{XMM-Newton}}
\begin{tabular}{l l l l l}
\hline\hline
Regions & Right Ascension & Declination & Radius \\
& h~m~s (J2000) & d m s (J2000) & & \\
\hline
Region 1 & 18:41:12.979 & $-$04:56:15.32 & 22$\arcsec$$\times$70$\arcsec$ \\
Region 2 & 18:41:15.744 & $-$04:56:00.56 & 22$\arcsec$$\times$60$\arcsec$ \\
Region 3 & 18:41:18.114 & $-$04:57:20.26 & 70$\arcsec$$\times$23$\arcsec$ \\
Region 4 & 18:41:21.275 & $-$04:57:52.74 & 84$\arcsec$$\times$23$\arcsec$ \\
Region 5 & 18:41:26.082 & $-$04:56:13.35 & 21$\arcsec$$\times$85$\arcsec$ \\
Region 6 & 18:41:23.514 & $-$04:55:49.73 & 15$\arcsec$$\times$36$\arcsec$ \\
Region 7 & 18:41:21.868 & $-$04:56:26.14 & 26$\arcsec$$\times$15$\arcsec$ \\
Region 8 & 18:41:19.695 & $-$04:55:33.99 & 40$\arcsec$$\times$18$\arcsec$ \\
Region 9 & 18:41:20.880 & $-$04:54:50.70 & 80$\arcsec$$\times$25$\arcsec$ \\
SNR & 18:41:19.343 & $-$04:56:11.16 & 150$\arcsec$\\
 \hline
\end{tabular}
\tablecomments{The  global SNR region is extracted from  a circular region of radius 150$\arcsec$ minus the emission from pulsar and other point sources, with the background extending from 150$\arcsec$--180$\arcsec$. \\
}
\end{table}

From both the \textit{Chandra} and \textit{XMM-Newton} observations, we defined 9 regions to characterize the diffuse emission from within the remnant.  The coordinates and the sizes of the extracted regions are summarized in Table~2 and shown in Figure~2. The source regions, marked by the green ellipses, were selected in such a way that the entire diffuse emission from within the remnant could be explored thoroughly. The western side of the SNR is covered by regions~1 and 2, southern side by regions~3 and 4, eastern side by region~5, and the northern part by regions~8 and 9. Regions~6 and 7 lie more toward the interior of the remnant. For the elliptical regions~1--9, the background was selected from a nearby source-free region to the northeast of the SNR on the \textit{Chandra} and \textit{XMM-Newton} chips and is shown by the cyan-colored dashed ellipse in Figure~2.
We further note here that the spectral fits for all small-scale regions were explored with different backgrounds (northwest and southwest regions near the SNR, and the SNR annular ring) and binning, and the spectral parameters obtained were consistent with those shown in Table~3.
In order to study the global SNR properties, we extracted the whole SNR region of radius 150$\arcsec$ with the emission from the pulsar and other point sources removed. This is shown in Figure~2, where the global SNR encloses the area inside the dashed cyan-colored annulus and the background extends from 150$\arcsec$--180$\arcsec$.  For the \textit{XMM-Newton} data, to maximize the SNR signal while minimizing any contamination by the AXP's spectrum, we excluded a circle centered at the AXP with a radius of 25$\arcsec$. We have investigated any possible contamination of the SNR's spectrum by the AXP's spectrum (especially in the hard X-ray band where its emission dominates) and found that it is negligible. In particular, we checked the pulsar's spectrum out to a larger radius of $\sim$35$\arcsec$ (which encompasses more than 90\% of the encircled energy for a point source) and found that the additional AXP flux (within 25$\arcsec$--35$\arcsec$) is negligible in comparison to the SNR's flux, with a relative AXP to SNR hard (4--10 keV) X-ray flux of only $\sim$2\%, which is within the error on the estimated SNR flux. Furthermore, the small-scale inner SNR regions extracted closest to the pulsar are at a radius $>$35", implying no contamination of these regions' spectra by the AXP. 

All the extracted spectra were analyzed using the X-ray spectral fitting package, XSPEC\footnote{http://xspec.gsfc.nasa.gov.}  version 12.6.0, and were grouped by a minimum of 25 and 50 counts per bin for regions 1--9 and the global SNR, respectively. The errors quoted are at the 90$\%$ confidence level.   

The spectra extracted from different regions within the SNR clearly showed the presence of line features. Hence, we started analyzing the emission from the remnant using thermal models of varying abundances. These include collisional ionization equilibrium (CIE) models such as VMEKAL (Mewe et al. 1985; Liedahl et al. 1995) and VRAYMOND (Raymond \& Smith 1977) used for modeling SNRs with optically thin thermal plasma that has reached collisional ionization equilibrium, and NEI models such as VPSHOCK, VNEI, and VSEDOV (Borkowski et al. 2001) appropriate for modeling young SNRs in which the plasma is still being ionized. We find that the spectral analysis using CIE models did not yield acceptable fits (reduced chi-squared $\chi^2_{\nu}$ $\ge$ 2 where $\nu$ is the number of degrees of freedom) while the NEI models with varying metal abundances resulted in much better fits.  

VPSHOCK is a plane-parallel non-equilibrium ionization plasma model with variable abundances, characterized by a constant electron temperature and a range of ionization timescales from 0 to $\tau$ = $n_et$, where $n_e$ is the post-shock electron density and $t$ is the time since the passage of the shock. The VNEI model also assumes variable abundances, but is characterized by a constant temperature and a single ionization parameter. VSEDOV is another NEI model with variable abundances based on the Sedov self-similar dynamics (Sedov 1959). This model is characterized by mean and electron temperatures immediately behind the shock and by the ionization timescale. 

\begin{table*}[th]
\caption{Best fit spectral parameters of the SNR Kes~73 using \textit{Chandra} and \textit{XMM-Newton}}
%\scriptsize
\begin{tabular}{l l l l l l l l l l l l}
\hline\hline Parameter & Reg 1 & Reg 2 & Reg 3 & Reg 4 & Reg 5 & Reg 6 & Reg 7 & Reg 8 & Reg 9 & SNR \\
\hline\hline

 $ N_{H}$ ($10^{22}$ cm$^{-2}$) & 3.3$^{+0.1}_{-0.3}$ & 3.2$^{+0.1}_{-0.7}$ & 2.8$^{+0.7}_{-0.5}$ & 2.7$^{+0.6}_{-0.3}$ & 2.9$^{+0.1}_{-0.5}$ & 2.6$^{+0.3}_{-0.7}$ & 2.8$^{+0.4}_{-0.5}$ & 2.9$^{+0.4}_{-0.2}$ & 2.7$^{+0.3}_{-0.8}$ & 2.6$^{+0.4}_{-0.3}$ \\

\cline{1-1}
Hard component& \\
\cline{1-1}

 $ kT_h$ (keV)& 1.1$^{+0.2}_{-0.1}$ & 1.4$^{+0.8}_{-0.1}$ & 1.7$^{+0.8}_{-0.2}$ & 1.7$^{+0.8}_{-0.2}$ & 1.1$^{+0.6}_{-0.1}$ & 1.3$^{+0.1}_{-0.3}$ & 1.6$^{+0.3}_{-0.2}$ & 1.7$^{+0.4}_{-0.3}$ & 1.3$^{+0.5}_{-0.1}$ & 1.6$^{+0.8}_{-0.7}$ \\

 Fe$_h$ &  1.5$^{+0.7}_{-0.9}$ & 2.3$^{+1.4}_{-1.5}$ &  1.6$^{+4.1}_{-0.7}$ & 2.4$^{+2.9}_{-1.0}$ & 0.9$^{+1.0}_{-0.5}$ & 2.0$^{+4.1}_{-1.1}$ & 1.2$^{+1.4}_{-0.8}$ &  1.2$^{+1.7}_{-1.0}$ & 1.1$^{+2.2}_{-0.8}$ &  1.4$^{+0.3}_{-0.4}$ \\

$n_et_h$ (10$^{11}$ cm$^{-3}$s) & 2.8$^{+1.1}_{-0.9}$ & 1.4$^{+0.7}_{-0.5}$ & 0.7$^{+0.3}_{-0.4}$ & 0.5$^{+0.1}_{-0.3}$ & 2.7$^{+1.4}_{-0.9}$ &  3.2$^{+2.0}_{-1.2}$ & 1.0$^{+0.5}_{-0.3}$ &  0.8$^{+0.9}_{-0.3}$ & 1.5$^{+0.5}_{-0.6}$ &  0.8$^{+0.2}_{-0.1}$ \\ 
 
$F_{unabs}$ (10$^{-11}$) &  9.8$^{+0.3}_{-2.0}$ & 6.6$^{+2.4}_{-4.3}$ & 4.3$^{+1.7}_{-3.0}$ & 3.2$^{+0.7}_{-1.8}$ & 3.7$^{+2.2}_{-1.2}$ & 1.7$^{+4.5}_{-0.6}$ & 2.4$^{+1.3}_{-0.7}$ &  3.6$^{+2.1}_{-0.8}$ &  2.8$^{+2.8}_{-1.1}$ & 181$^{+16}_{-45}$ \\

\cline{1-1}
Soft component & \\
\cline{1-1}

$kT_s$ (keV)& 0.3$^{+0.1}_{-0.1}$ &0.4$^{+0.2}_{-0.1}$ &  0.5$^{+0.1}_{-0.3}$ & 0.5$^{+0.2}_{-0.1}$ & 0.3$^{+0.2}_{-0.1}$ & 0.5$^{+0.2}_{-0.2}$ & 0.4$^{+0.1}_{-0.1}$ &  0.4$^{+0.1}_{-0.1}$ & 0.5$^{+0.1}_{-0.2}$ & 0.5$^{+0.1}_{-0.2}$ \\

O & 1.6$^{+4.1}_{-0.9}$ & 1.9$^{+3.9}_{-1.3}$ & 1.1$^{+0.8}_{-0.6}$ & 1.2$^{+0.7}_{-0.6}$ &  1.4$^{+1.0}_{-1.4}$ & 1.6$^{+0.6}_{-0.7}$ & 1.0$^{+3.9}_{-1.0}$ &  1.0$^{+3.3}_{-0.5}$ & 1.0$^{+1.8}_{-0.7}$ & 1.8$^{+1.2}_{-0.4}$ \\

Mg & 1.7$^{+2.5}_{-0.7}$ & 0.9$^{+1.6}_{-0.6}$ & 0.9$^{+1.2}_{-0.3}$ & 0.8$^{+0.5}_{-0.8}$ &  1.2$^{+1.0}_{-1.2}$ & 1.2$^{+1.1}_{-1.2}$ & 0.5$^{+0.2}_{-0.1}$ &  0.6$^{+1.4}_{-0.4}$ &  1.0$^{+0.6}_{-0.8}$ & 1.0$^{+1.4}_{-0.7}$  \\

Si & 5.9$^{+2.5}_{-0.7}$ &  2.3$^{+0.8}_{-0.5}$ & 1.4$^{+5.6}_{-0.3}$ & 1.6$^{+4.1}_{-0.5}$ &  5.3$^{+3.5}_{-1.4}$ & 2.8$^{+3.1}_{-1.6}$ & 1.1$^{+0.5}_{-0.5}$ &  1.4$^{+0.9}_{-0.6}$ & 1.9$^{+3.2}_{-0.2}$ & 1.7$^{+2.0}_{-0.5}$ \\

S & 13.2$^{+5.7}_{-7.2}$ &  4.3$^{+1.7}_{-1.1}$ & 2.5$^{+2.3}_{-0.8}$ & 2.8$^{+3.1}_{-1.3}$ & 10.2$^{+4.2}_{-7.2}$ & 4.1$^{+4.5}_{-2.2}$ & 2.6$^{+2.0}_{-0.8}$ & 3.5$^{+3.8}_{-1.8}$ &  3.0$^{+7.4}_{-1.0}$ & 2.9$^{+1.3}_{-1.6}$ \\

Fe$_s$ & 1.5$^{+0.7}_{-0.9}$ & 0.6$^{+0.7}_{-0.4}$ & 0.4$^{+1.0}_{-0.4}$ & 0.3$^{+0.4}_{-0.3}$ & 1.0$^{+2.5}_{-1.0}$ &  0.6$^{+1.4}_{-0.3}$ &  0.2$^{+0.5}_{-0.2}$ & 0.7$^{+1.3}_{-0.7}$ &  0.6$^{+0.8}_{-0.2}$ & 0.3$^{+0.2}_{-0.2}$  \\

 $n_et_s$ (10$^{12}$ cm$^{-3}$s) & $>$1.2 & 0.5$^{+0.6}_{-0.2}$ & $>$ 4.4 & $>$1.8 & $>$ 0.6 & $>$1.0 &  0.4$^{+0.6}_{-0.1}$ & $>$0.8 & $>$0.7 & $>$6.1 \\

$F_{unabs}$ (10$^{-10}$) & 2.3$^{+1.4}_{-0.7}$ & 3.4$^{+0.7}_{-0.9}$ & 0.9$^{+0.6}_{-0.8}$ & 4.4$^{+0.8}_{-0.9}$ &  1.1$^{+0.4}_{-0.3}$ &  3.7$^{+2.3}_{-0.3}$ & 2.2$^{+1.1}_{-0.6}$ &  1.0$^{+3.0}_{-1.2}$ & 0.5$^{+0.8}_{-0.2}$ &  20.6$^{+7.9}_{-3.7}$\\

\hline

$F_{abs}$ (10$^{-11}$) & 0.4$^{+0.1}_{-0.1}$ & 0.3$^{+0.1}_{-0.1}$ & 0.4$^{+0.2}_{-0.1}$ & 0.2$^{+0.2}_{-0.3}$ &  0.2$^{+0.2}_{-0.2}$ &  0.2$^{+0.2}_{-0.1}$ & 0.2$^{+0.2}_{-0.3}$ &  0.2$^{+0.1}_{-0.2}$ & 0.2$^{+0.1}_{-0.3}$ &  12.8$^{+3.2}_{-2.1}$\\

 $\chi_{\nu}^2/dof$ & 1.29/708 &  1.11/652 & 1.29/753 & 1.14/605 & 1.15/516 & 1.26/372 & 1.29/375 &  1.32/410 & 1.06/504 & 1.51/1743 \\

\hline
\end{tabular}
\tablecomments{Regions~1--9 and the global SNR spectra are fitted using the VPSHOCK+VPSHOCK and VSEDOV+VPSHOCK models, respectively. All elemental abundances are in solar units given by Anders \& Grevesse (1989). The subscripts `$s$' and `$h$' denote soft and hard components, respectively. The 0.5--10.0 keV unabsorbed ($F_{unabs}$) and total absorbed ($F_{abs}$) fluxes quoted are in units of ergs~cm$^{-2}$~s$^{-1}$. The given uncertainties on each fitted value are at the 90\% confidence level for a single interesting parameter. }  
\end{table*}
\begin{figure*}[h]
\includegraphics[width=0.5\textwidth]{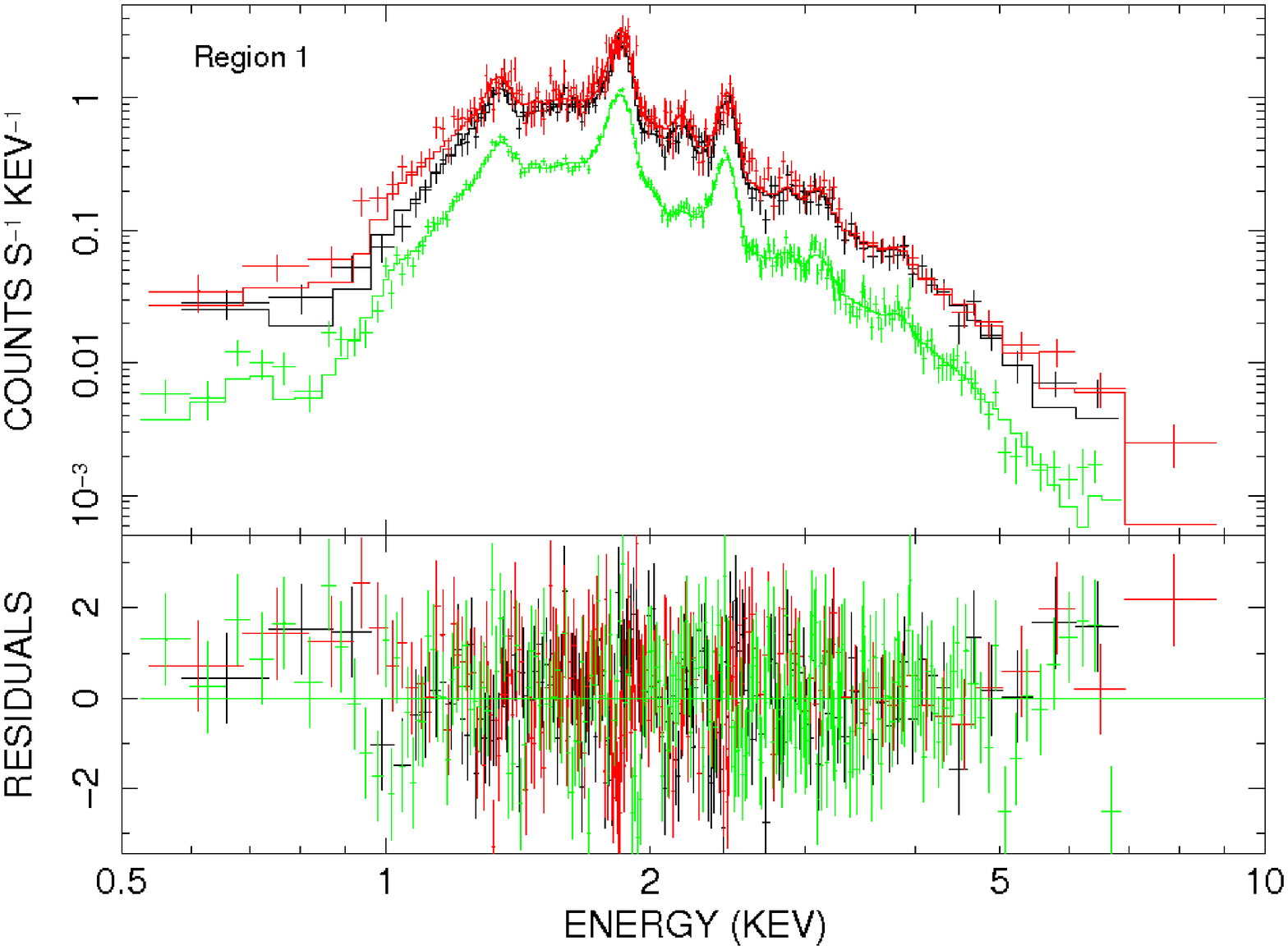} 
\includegraphics[width=0.5\textwidth]{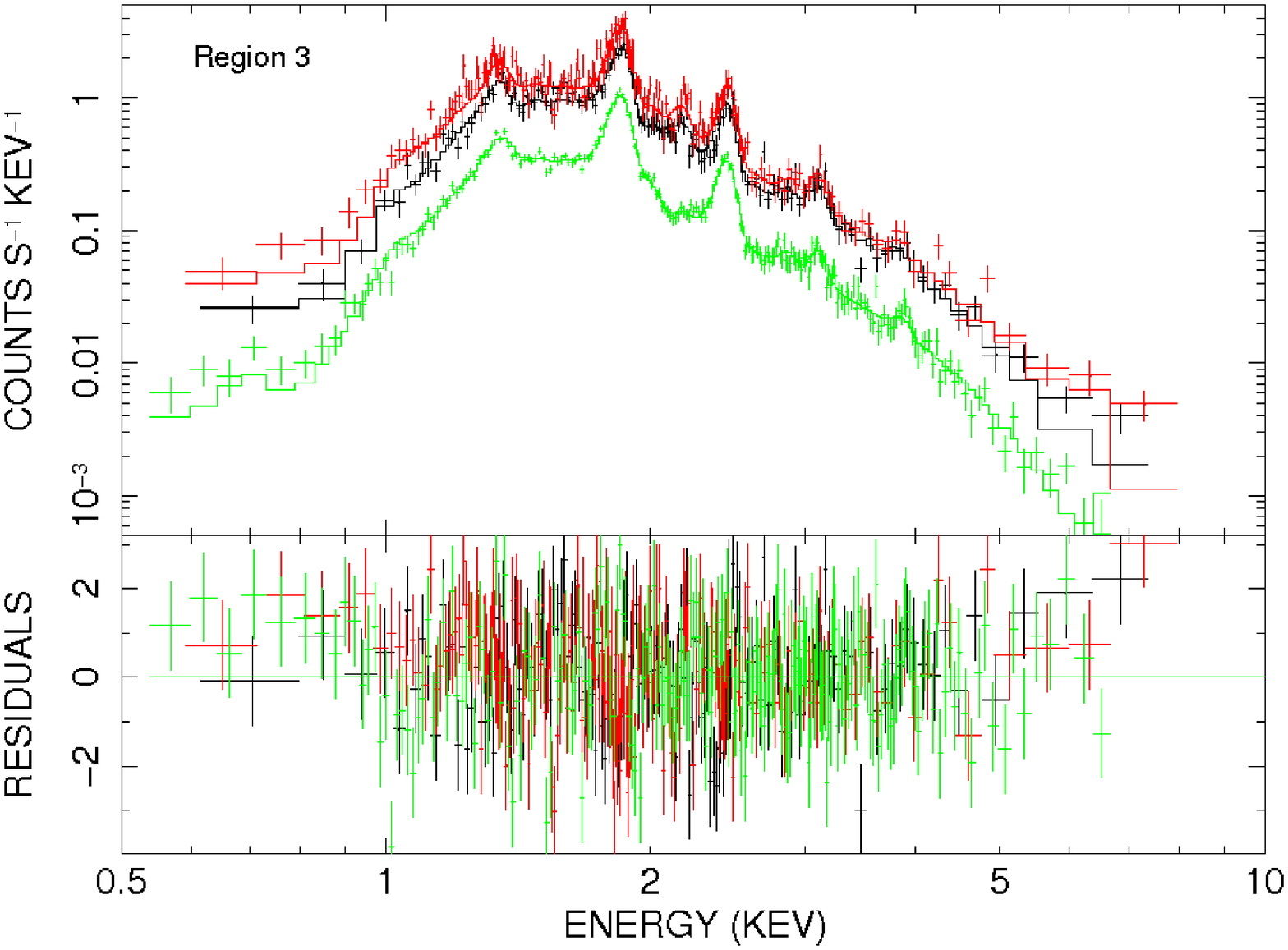} 
\includegraphics[width=0.5\textwidth]{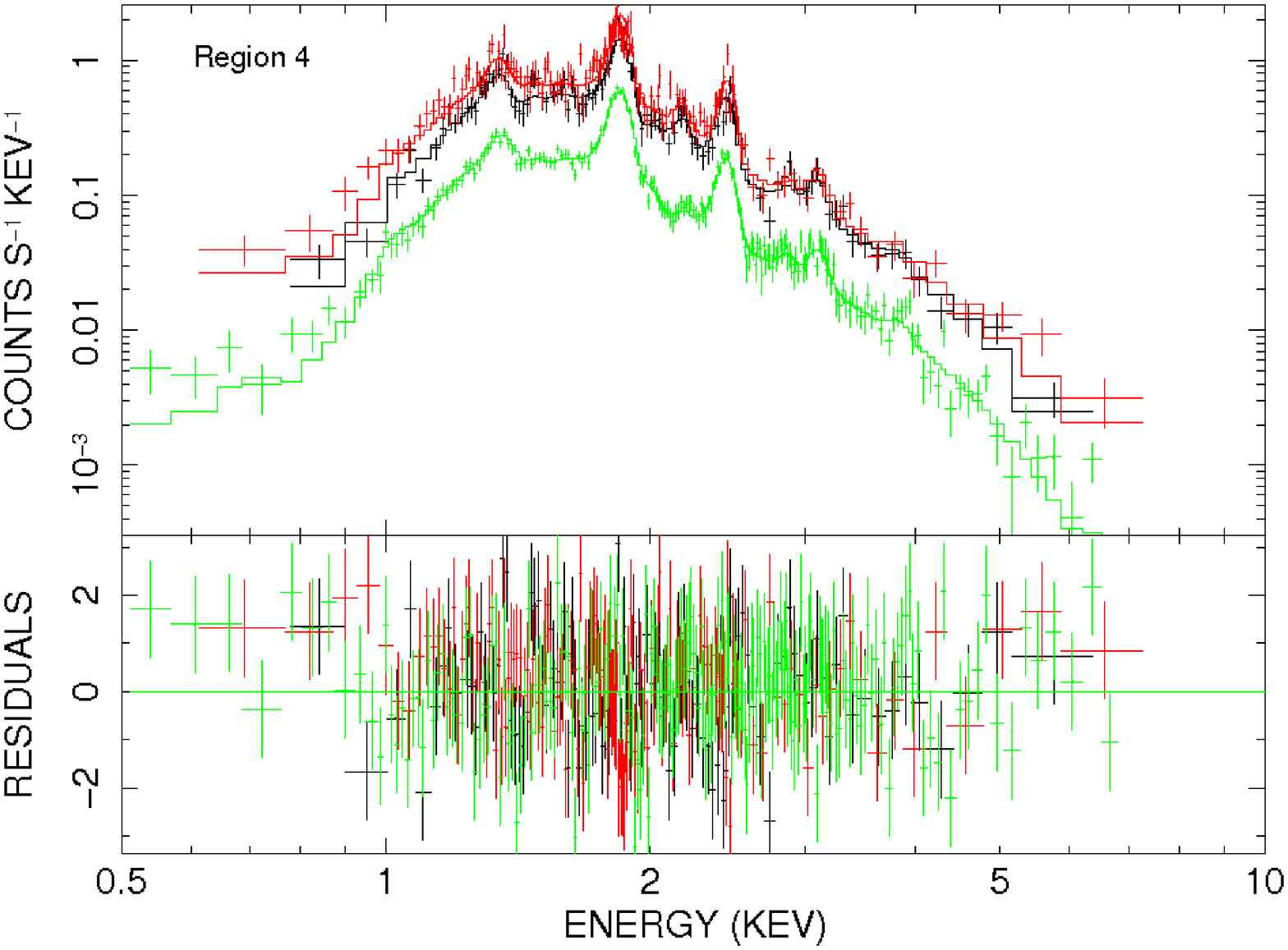} 
\includegraphics[width=0.5\textwidth]{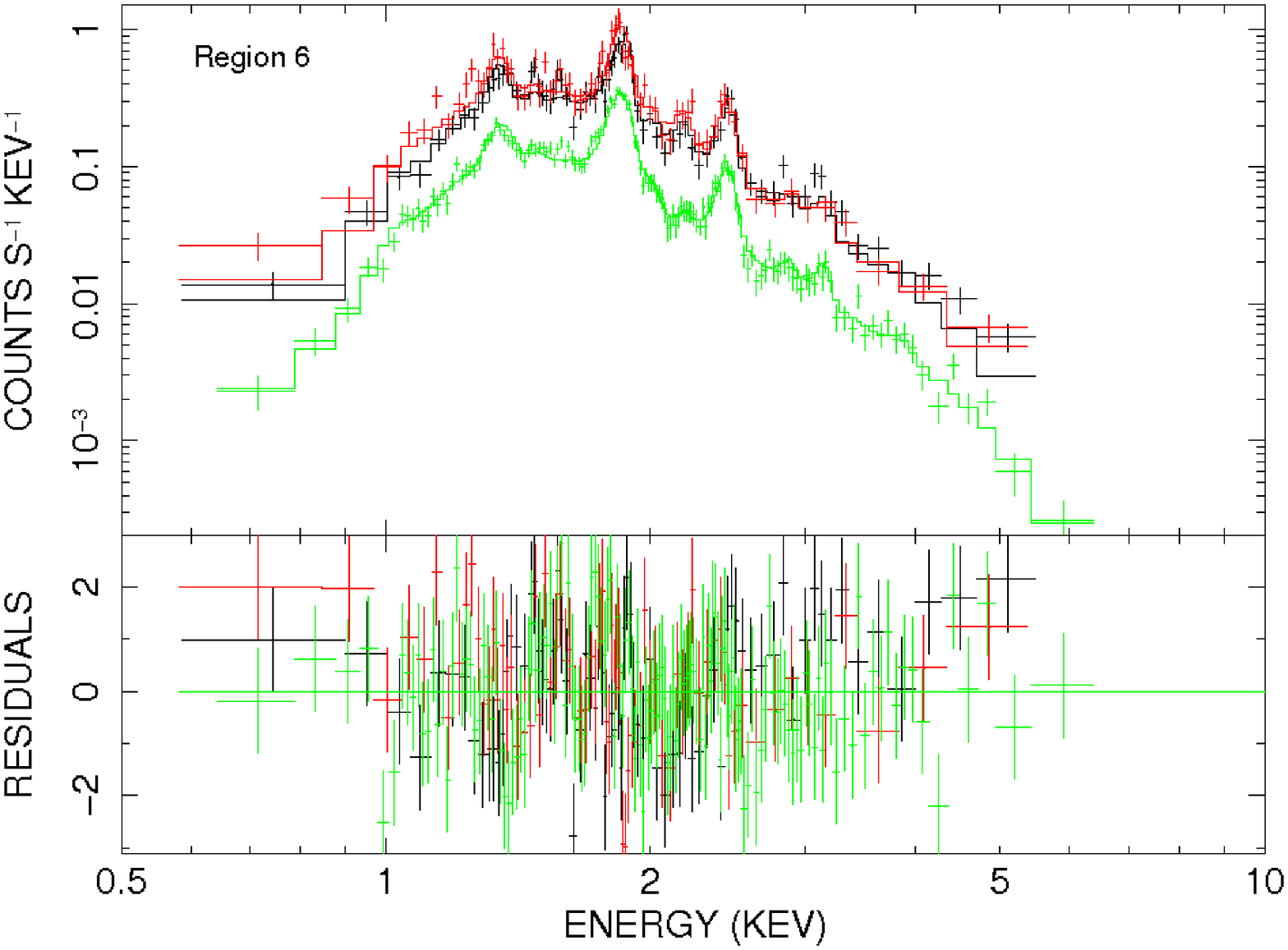} 
\includegraphics[width=0.5\textwidth]{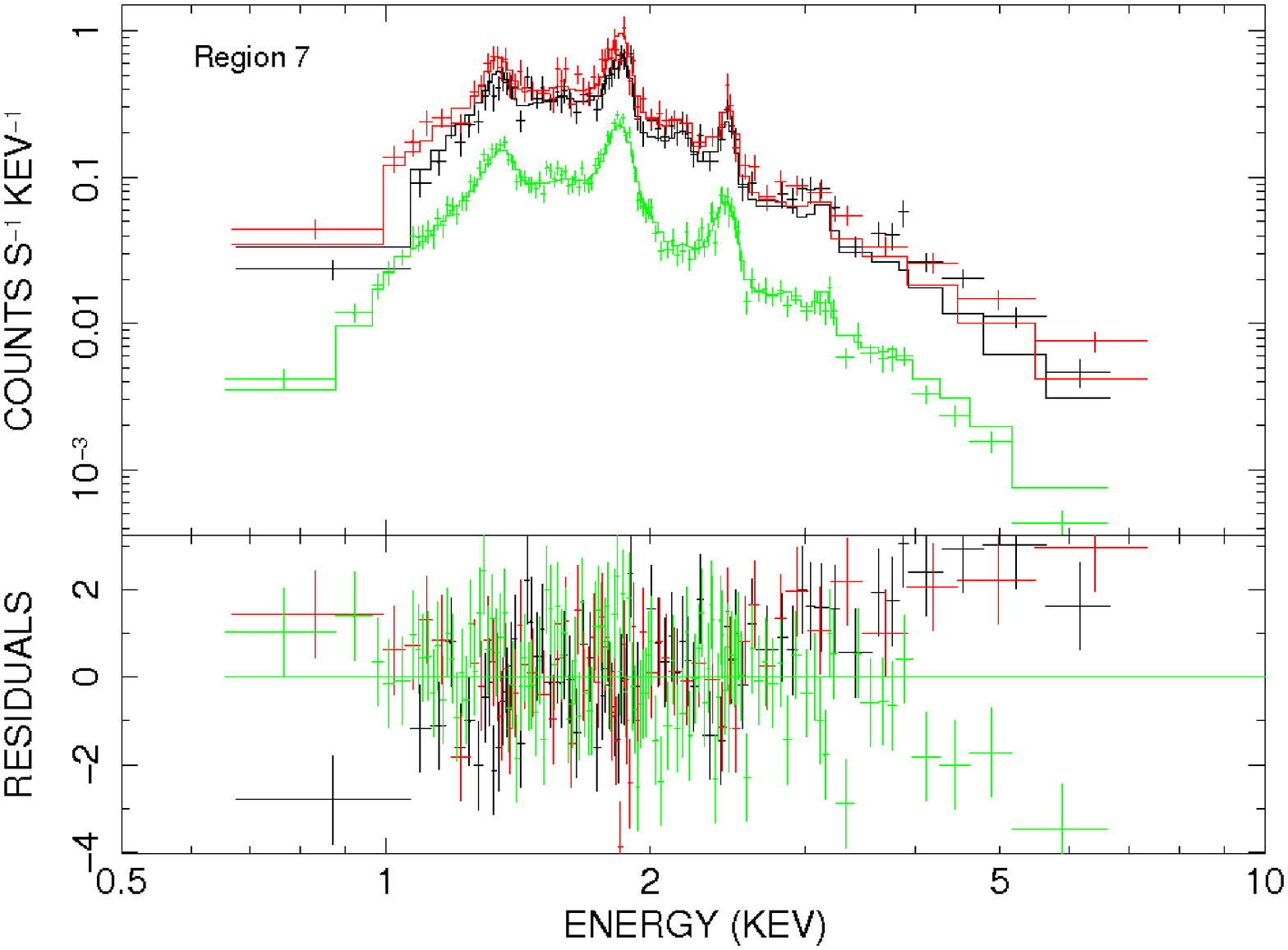} 
\includegraphics[width=0.5\textwidth]{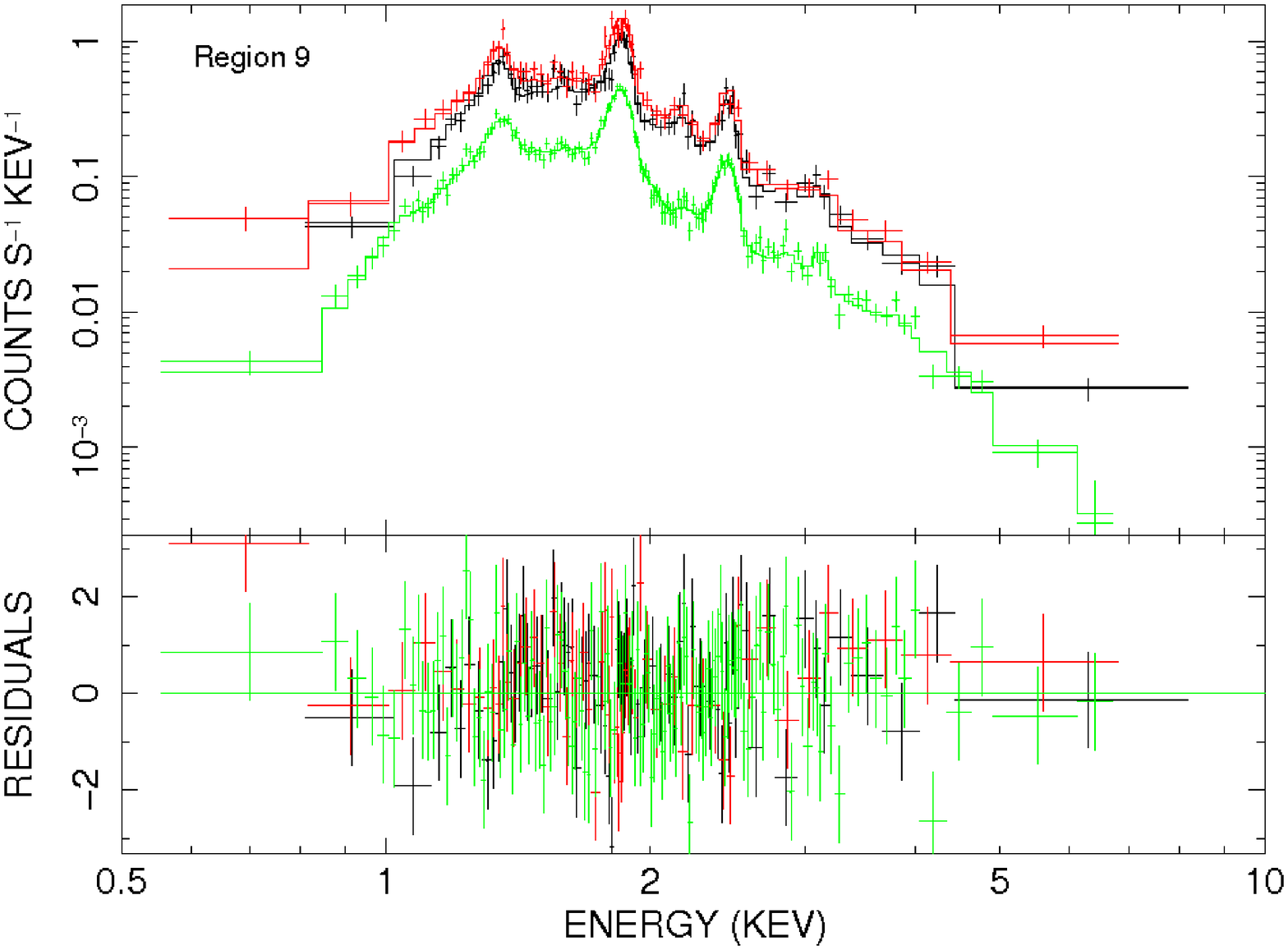} 
\caption{Examples of VPSHOCK+VPSHOCK model fit to selected regions. The spectra shown refer to the combined (and rebinned) \textit{XMM-Newton} pn (red), MOS1+2 (black), and \textit{Chandra} (green) data for display purposes. Representative regions chosen are the bright western limb (region~1), the bright SNR interior (regions~3, 6, and 7), the shell-like feature on the southern side (region~4), and the faint northern side (region~9).  }
\end{figure*}

\begin{figure*}[htbp]
\includegraphics[width=\textwidth]{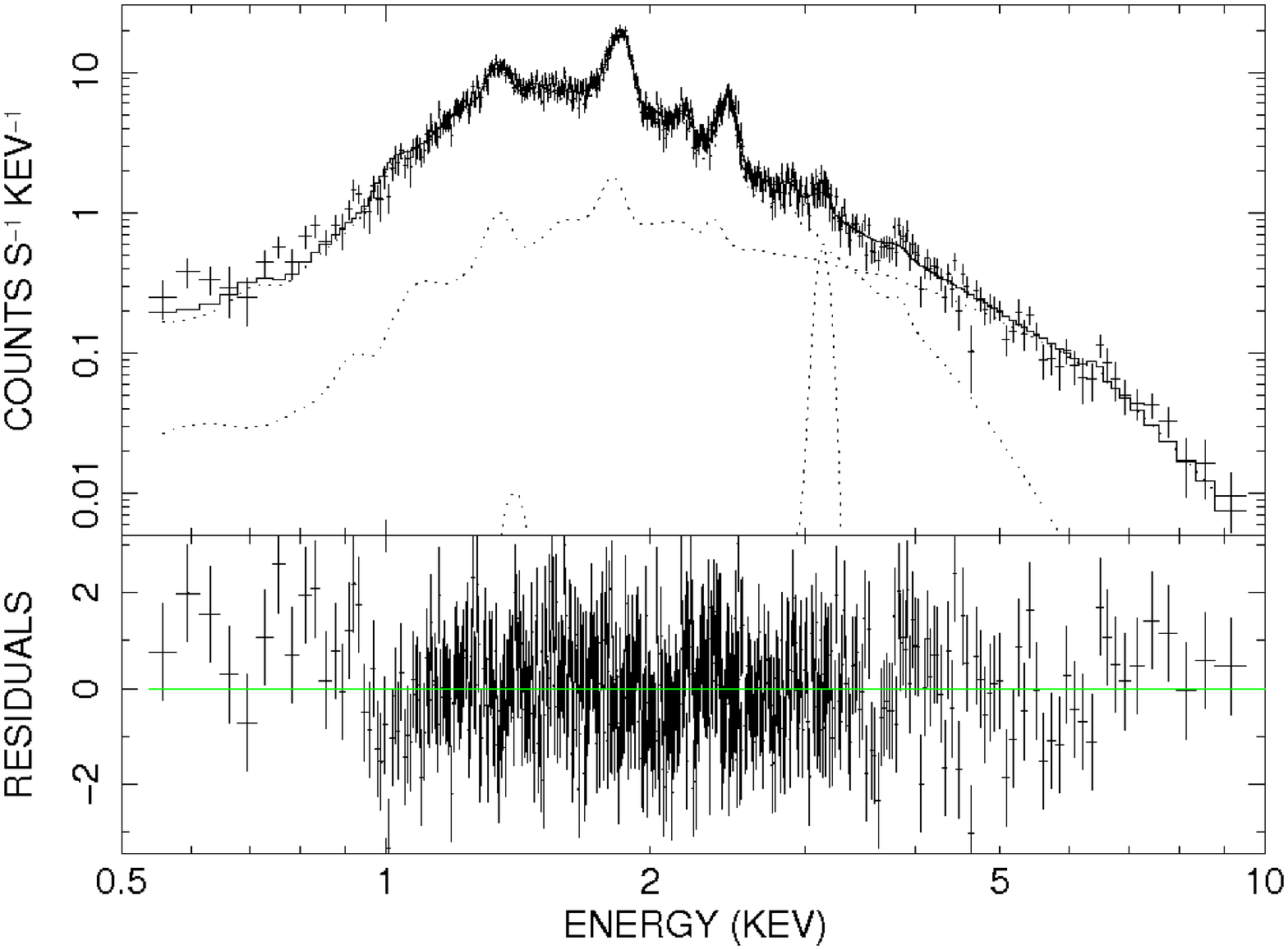}
\caption{ \textit{XMM-Newton} pn spectrum extracted from the whole SNR, fitted with a two-component thermal model (VSEDOV+VPSHOCK). A Gaussian line around 3.1~keV is also added to account for emission from Argon. The pn spectra from the two \textit{XMM-Newton} observations have been combined together for display purposes here. The dotted lines represent the contributions of the soft and hard model components, along with the contribution from the Argon line emission added to the model. }
\end{figure*}

\subsection{Global fitting and spatially resolved spectroscopy}
\label{4.1}

The \textit{Chandra} and \textit{XMM-Newton} spectra of the SNR regions were first investigated with single component VPSHOCK models, but none of them gave a satisfactory fit ($\chi^2_{\nu}$ $\ge$ 2) and did not account for residuals above $\sim$4 keV unlike previous studies with poorer statistics (Gotthelf \& Vasisht 1997; Vink \& Kuiper 2006). Since the X-ray spectrum of a young SNR is expected to show distinct components associated with the swept-up ISM and ejecta, the spectra were subsequently examined using two-component thermal models.  The addition of a second thermal component (e.g., a VPSHOCK+VPSHOCK model) significantly improved the fits for all the regions.  Throughout the paper, we use the subscripts `$s$' and `$h$' to denote the soft and hard components, respectively. 

The abundances of all the elements were initially fixed at their solar values given by Anders \& Grevesse (1989).  All the models included the Morrison \& McCammon (1983) interstellar absorption\footnote{The fits were also explored with the Tuebingen-Boulder ISM absorption model ($tbabs$ in $XSPEC$) and the spectral parameters were consistent with those given in Table 3.} and Ni was tied to Fe throughout.  A Gaussian line near 3.1~keV was added to the two-component models since the VPSHOCK model used in XSPEC does not account for the Ar emission line visible in the spectra. The spectra were first fit by treating the column density, temperature, and normalization as free parameters and then letting the abundances vary freely as needed. We obtained nearly solar values for all the hard component elements except for a marginally enhanced Fe, while the soft component showed enhanced abundances for Mg, Si, and S, suggesting that the soft component is associated with shock-heated ejecta. Allowing Fe$_h$ to vary improved the $\chi^2_{\nu}$ value with an $F$-test probability $\lesssim$10$^{-4}$. The emission lines from O lie in the soft X-ray band ($<$0.6~keV) and are hard to detect due to the large column density of the SNR ($N_H$ $\sim$ 2.6$\times$10$^{22}$ cm$^{-2}$). However, when allowed to vary, the O abundance improved the spectral fits with an $F$-test probability of $\sim$10$^{-3}$ for some regions and for the global SNR's spectrum. We also note a slightly enhanced Ca abundances for some regions, but allowing Ca to vary did not improve the fits significantly and posed some difficulty in constraining the errors. Therefore, the Ca abundance was frozen to its solar value for the small scale regions during spectral fitting.  In  summary, the spectral fitting for all regions was performed by leaving the hard component Fe abundance (Fe$_h$) and the soft component O, Mg, Si, S, and Fe$_s$ abundances as free parameters\footnote{Even with O frozen to solar, the abundance ratios with respect to Si (see Section 5.4) and the spectral parameters obtained were consistent with the tabulated values within error.}.Table~3 summarizes the best fit spectral parameters obtained with a VPSHOCK+VPSHOCK model\footnote{We also performed the spectral fitting using a VNEI+VNEI model, which yielded values consistent with those obtained with a VPSHOCK+VPSHOCK model.} for all regions, and  Figure~3 shows the spectra of some of the SNR regions of interest: the bright western limb (region~1), bright interior regions~3, 6, and 7, the shell-like feature on the southern side (region~4), and the relatively faint northern edge (region~9).

For the global SNR spectrum, we performed spectral fitting using the VSEDOV+VPSHOCK model plus a Gaussian line near 3.1~keV (Table~3).  Here, the VSEDOV model is associated with the hard component and yielded the same electron and ion temperatures, while the VPSHOCK model is associated with the soft component.  The SNR spectrum was further explored using a VPSHOCK+VPSHOCK model and the spectral parameters were found to be in good agreement with those obtained using the VSEDOV+VPSHOCK model (Table~3), with a slightly different abundance values for Si = 2.3$^{+3.3}_{-1.1}$, S = 4.2$^{+4.6}_{-0.9}$, and Fe$_s$ = 0.6$^{+0.8}_{-0.2}$.  In Figure~4, we show the \textit{XMM-Newton} pn spectrum extracted from the whole SNR fitted with a VSEDOV+VPSHOCK model, displaying the different model contributions as dotted lines.

In order to check for any underlying synchrotron component originating from the acceleration of particles at the supernova shock, we explored the spectra of all extracted regions by adding a non-thermal model (e.g., power-law) to the one- and two-components thermal models. The addition of a power-law component to the one-component thermal models (VPSHOCK or VNEI) resulted in a large $\chi^2_{\nu}$ $\ge$ 1.8 value for all regions, confirming that the second component is better described by a thermal model. When added to the best fit two-component thermal models (VPSHOCK+VPSHOCK or VNEI+VNEI), the additional power-law component slightly improved the fits for some of the regions. For the global SNR spectrum, the additional power-law component added to the best-fit VSEDOV+VPSHOCK model yielded an F-test probability of $\geq$10$^{-4}$ with a photon index $\Gamma$ = 3.4$^{+1.8}_{-1.5}$ and an upper limit on the non-thermal luminosity of $\sim$5$\times$10$^{32}$~erg~s$^{-1}$ (at an assumed distance of 8.5~kpc). Adding a power-law component to the best fit VPSHOCK+VPSHOCK model for the global SNR spectrum also resulted in similar spectral parameters. We note a caveat for fitting a synchrotron component with a simple power-law model. More appropriate models include the $srcut$ model in XSPEC; however due to uncertainties on the radio properties of our selected regions and the marginal need for an additional non-thermal component, we did not explore this further. A more detailed radio and X-ray spatially resolved study would require deeper observations.

As seen from Table~3, there is no significant variation in the spectral parameters across the remnant, except for the column density ($N_H$) which appears slightly elevated on the western side for region~1. The estimated difference in $N_H$ between the average SNR and the western side\footnote{We note here that the $N_H$ value, although model dependent, was not affected by varying the O abundance, as discussed above.} is $\sim$7$\times$10$^{21}$~cm$^{-2}$.  To the best of our knowledge, there is no evidence of any observed molecular cloud towards the western side of the remnant that might account for the higher absorption. Tian \& Leahy (2008) suggest the existence of a CO cloud (at 89 $\pm$ 2 km~s$^{-1}$) in the direction of Kes~73, but located behind the remnant and therefore, the cloud is not absorbing the SNR. Moreover, we explored the environment of Kes~73 in the radio and infrared wavelengths (discussed in Section 5.1) and do not find any evidence of a cloud or nearby sources that appear to interact with the SNR.  This is consistent with the overall circular morphology of the SNR.

We note that Lopez et al. (2011) analyzed spectra from a large number of small scale regions ($\sim$15$^{\prime\prime}$ radius), and reported good fits for one-component models using data from only the earliest \textit{Chandra} observation of Kes~73 (ObsID~729), providing evidence for enhanced elemental abundances across the SNR and an elevated value of $N_H$ for the western side. We obtain similar results using all of the available \textit{Chandra} observations of Kes~73, and using somewhat larger spectral regions, but find that two-component thermal models are required for all of our extracted regions. To check this apparent discrepancy, we extracted spectra from a representative $\sim$15$^{\prime\prime}$ radius region using only data from ObsID~729. We found that a single-component VPSHOCK adequately describes the spectrum, and that the improvement in the fit from addition of another VPSHOCK model is not statistically significant (F-test probability of $\sim$0.05).  On the other hand, for slightly larger region sizes (e.g., region 7 in Figure~2) using the same single data set, the addition of a second VPSHOCK model is required to adequately fit the spectrum, with an F-test chance probability of $\sim$6$\times$10$^{-6}$. In summary, the differences in spectral results between our work and that of Lopez et al. (2011) are primarily due to the larger region sizes and the inclusion of all available data for Kes~73.

\section{Discussion}
\label{5}

\subsection{Multi-wavelength morphology and environment of SNR Kes~73}
\label{5.1}

\begin{figure*}[th]
\vspace{-0.8cm}
\center
\includegraphics[width=0.8\textwidth]{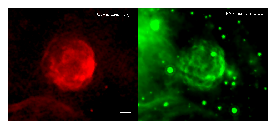} 
\includegraphics[width=0.82\textwidth]{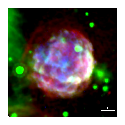}
\caption{\textit{Top}: MAGPIS 20 cm radio (left) and MIPSGAL 24$\mu$m infrared (right) images of the SNR Kes~73. The radio and IR images have been smoothed with a Gaussian with $\sigma$~=~3$\arcsec$ and have a field of view of 10$\arcmin$.5$\times$10$\arcmin$.5 with the AXP at the center. The foreground stars appear as saturated point sources (in green) in the infrared image. \textit{Bottom}: Tri-color image of SNR Kes~73 with the radio emission shown in red, infrared emission in green, and X-ray emission in blue. North is up and East is to the left.}
\end{figure*}

We investigate the multi-wavelength morphology of Kes~73 using the radio, infrared, and X-ray data. The radio image was extracted from the Multi-Array Galactic Plane Imaging Survey (MAGPIS\footnote{http://third.ucllnl.org/gps/}) at a wavelength of 20 cm for the region 5$^{\circ}$ $<$ $l$ $<$ 48$^{\circ}$.5, $|b|$ $<$ 0$^{\circ}$.8 (Helfand et al. 2006) and the infrared image from the survey of the inner Galactic plane using \textit{Spitzer}'s Multiband Imaging Photometer (MIPS) at a wavelength of 24 $\mu$m (MIPSGAL\footnote{http://mipsgal.ipac.caltech.edu}) for the region 4$^{\circ}$ $<$ $l$ $<$ 50$^{\circ}$, $|b|$ $<$ 1$^{\circ}$.0 (Carey et al. 2009). In Figure~5 (top), we show the radio (left) and infrared (right) images of the SNR, with a field of view of $\sim$10$\arcmin$.5$\times$10$\arcmin$.5 centered around the pulsar. We estimate an apparent diameter of $\sim$5$\arcmin$ for the remnant in radio and infrared wavelengths in comparison with the slightly smaller diameter of $\sim$4$\arcmin$.5 in X-rays. We have also constructed a composite color image of the radio (red), infrared (green), and X-ray (blue) emission of SNR Kes~73 as shown in Figure~5 (bottom).  The X-ray image shown here corresponds to the broadband (0.5--7.0 keV) \textit{Chandra} image, adaptively smoothed (using the \textit{csmooth} command in CIAO) using a Gaussian kernel of $\sigma$~=~1$\arcsec$--2$\arcsec$ and for a significance of detection 2 to 5.
 
As mentioned earlier, Kes~73 has been classified as a shell-type remnant in the radio (Kriss et al. 1985). Now, we compare the radio and X-ray morphologies of the remnant and observe a few similarities. The remnant exhibits a spherical morphology with the western limb (which has the highest $N_H$ and lowest $kT_h$) appearing the brightest in both wavelengths, possibly due to some local density enhancements in that region.  The radio shell extending from the eastern limb to the southern edge of the SNR matches with the corresponding quarter shell-like structure (i.e., regions~4 and 5 in Figure~2) observed in the X-ray image, thus appearing purple in Figure~5 (bottom). The enhanced Si and S abundances obtained from the spectral fits (Table~3) also suggest the presence of reverse shocked ejecta in these regions.  We also notice some faint radio clumps on the northern side of the SNR spatially correlating with the X-ray image.  Apart from these few similarities, the remnant's morphology also differs in these wavelengths with the apparent lack of bright radio emission from the SNR interior and the pulsar (as expected due to the nature of their emission).  The radio shell appears more extended on the northeastern side of the SNR and displays a thin filamentary structure along the edges in pure red (Figure~5). This feature is probably the expanding outer blast wave, which is not detected in X-rays. In summary, we see more diffuse X-ray emission originating from the center of the SNR due to the reverse shocked ejecta and filling the radio shell, as seen from the blue color at the SNR interior in Figure~5 (bottom). 

The \textit{Spitzer} MIPS image (Figure~5, top right) reveals a morphology more comparable to that of the X-ray image with arc-like features and bright infrared emission filling the SNR. The western limb of the SNR is very bright in the infrared as well, precisely outlining the radio and X-ray emission, and so appears white in the multi-wavelength image of Kes~73. The interior structures partially coincide with the X-ray image, but no significant infrared emission is seen from the northeastern boundary of the SNR. The infrared image shows that the remnant lies in a complex region of the Galactic plane, as previously noted by other authors (e.g., Wachter et al. 2007), with several saturated foreground or background sources, appearing in green. Pinheiro Goncalves et al. (2011) also suggested a strong morphological association of the 24 $\mu$m infrared emission and X-ray emission in SNR Kes~73, although it is not clear whether the strong 24 $\mu$m emission is the result of line emission or hot dust. 

\subsection{Blast wave and ejecta distribution}
\label{5.2}

We discuss here the origin of the soft and hard components derived from  our spectral fits. Our spatially resolved spectroscopic study confirms the need for a two-component thermal model to describe the X-ray emission originating from the small scale diffuse emission regions.  As discussed in Section~4.2, the soft component, with plasma temperatures $\sim$0.3--0.5~keV, shows enhanced metal abundances while the hard component, with plasma temperatures $\sim$1.1--1.7~keV, is characterized by solar abundances (except for a slightly higher Fe$_h$ for some regions). These results indicate that the soft component arises primarily from the reverse shocked ejecta and the hard component is dominated by emission from the forward shock. Furthermore, the soft component has reached ionization equilibrium ($n_et_s$ $\gtrsim$~10$^{12}$~cm$^{-3}$~s) for most of the regions as opposed to the under-ionized plasma ($n_et_h$ $\sim$ (0.5--3.2)$\times$10$^{11}$~cm$^{-3}$~s) characterizing the hard component. Interestingly, the results we obtained for Kes~73 are also very similar to those obtained for the mixed-morphology remnant 3C 397 (Safi-Harb et al. 2005). In particular, 3C 397 also needed a two-component model to fit the X-ray emission from all small scale regions, with the hard component showing a lower ionization timescale while the soft component was closer to ionization equilibrium. As well, the soft component was dominated by ejecta  requiring enhanced abundances while the hard component showed solar abundances except for Fe$_h$ and was interpreted as the presence of reverse shocked Fe bubbles (Safi-Harb et al. 2005). However, the abundance of Fe$_h$ in Kes~73 is only marginally enhanced unlike in 3C 397 whose spectrum is characterized by a very strong Fe-K line not detected in Kes~73.

As mentioned above, we detect enhanced metal abundances in the soft spectral component (see Table~3). S appears to be highly enhanced across the remnant followed by Si (for regions~1, 2, 5, 6, and 9). Mg and O seem slightly enhanced for regions~1 and 2, respectively. Fe$_s$ is seen enhanced only for region~1 while Fe$_h$ is slightly enhanced for regions~1, 2, 3, 4, and 6. Most notably, the western and eastern X-ray limbs of the remnant (regions~1 and 5) exhibit very similar spectral properties (although $N_H$ appears slightly higher for the brighter region~1) with elevated and comparable Si and S abundances. The emission from these regions correlates well with their radio counterpart and appears to fall within the outermost radio shell, the red emission seen along the outer edges in Figure~6 (bottom). These regions likely represent the reverse shocked ejecta seen in X-rays. The outermost fainter region seen in the radio image, which most likely corresponds to the forward shock, is not clearly detected in X-rays. In summary, we observe enhanced Si and S abundances for all regions across the remnant suggesting the presence of ejecta.

\subsection{X-ray properties of SNR Kes~73}
\label{5.3}

In the following, we estimate the physical properties of the diffuse emission regions across the remnant and the supernova explosion parameters using the spectral fit parameters obtained with the VPSHOCK/VSEDOV models summarized in Table~3. The derived quantities are shown in Table~4. Here, we take the distance to the SNR as 8.5 kpc (Tian \& Leahy 2008), noting that an earlier, slightly smaller, distance estimate of 6--7.5 kpc was obtained by Sanbonmatsu \& Helfand (1992). We introduce a scaling factor $d_{8.5}$ = $d$/8.5 kpc to account for the distance uncertainty and take a radius of 2$\arcmin$.5 for the SNR (as estimated from the extent of the radio emission, see Figure 5), which corresponds to a physical size $R_s$ = 6.2~$d_{8.5}$~pc = 1.9$\times$10$^{19}$~$d_{8.5}$~cm. 

The volume of the X-ray emitting regions ($V$) is estimated by assuming that the plasma fills an ellipsoid with the semi-major and semi-minor axes equivalent to those of the extracted SNR regions (Table~2) and the depth along the line of sight equal to the long axes of the extracted regions. 
We then introduce the volume filling factors for the cold ($f_s$) and hot ($f_h$) plasma and assume that $f_s$ + $f_h$ = 1. The emission measure for each component is related to the amount of plasma available to produce the observed flux and is given by $EM$ = $\int$$n_e n_H dV$ $\sim$ $f n_e n_H V$ where $n_e$ is the electron density of the shocked gas and $n_H$ is the proton density ($n_e$ = 1.2$n_H$ assuming cosmic abundances). By assuming pressure equilibrium between the soft and hard thermal components ($n_sT_s$ $\approx$ $n_hT_h$), the relative filling factors can then be derived as $f_s$ = [1 + $(EM_h/EM_s)(T_h/T_s)^2$]$^{-1}$; $f_h$ = 1~$-$~$f_s$ (see Table~4).

For cosmic abundances and strong shock Rankine-Hugoniot jump conditions, the ambient density $n_0$ can be estimated from the electron density $n_e$ as $n_e$ = 4.8$n_0$; here $n_0$ includes only hydrogen (Borkowski et al. 2001). Using the normalization value $K$ = (10$^{-14}$/4$\pi$d$^2)\int$$n_e n_H dV$ determined from the X-ray spectral fits in \textit{XSPEC}, we estimate the EMs of the two thermal components as well as the corresponding densities $n_e$ and $n_0$ (derived from $n_{e_h}$ of the hard component) of the X-ray emitting gas, as shown in Table~4. The electron densities inferred for the soft component are higher than those associated with the hard component. The ambient density, obtained from the global fit to the SNR using the VSEDOV component\footnote{We get $n_0$ $\sim$ 0.6~cm$^{-3}$ using the VPSHOCK+VPSHOCK model, consistent with the value obtained using the VSEDOV+VPSHOCK model.}, is $n_0$=~0.5$^{+0.4}_{-0.2}$~$f_h^{-1/2}$$d_{8.5}^{-1/2}$~cm$^{-3}$. The total unabsorbed flux of the SNR in the 0.5--10.0 keV energy range is $F_X$ = 3.9$^{+0.8}_{-0.6}$$\times$10$^{-9}$~ergs~cm$^{-2}$~s$^{-1}$, which corresponds to an X-ray luminosity of $L_X$ = 3.3$^{+0.7}_{-0.5}$$\times$10$^{37}$~$d^2_{8.5}$~ergs~s$^{-1}$. 

\begin{table*}[th]
\caption{Derived X-ray parameters of SNR Kes~73}
%\scriptsize
\begin{tabular}{l l l l l l l l l l l l}
\hline\hline
Parameters & Reg 1 & Reg 2 & Reg 3 & Reg 4 & Reg 5 & Reg 6 & Reg 7 & Reg 8 & Reg 9 & SNR\\
\hline\hline
\textit{Soft component} & \\
\cline{1-1}

 EM$_s$ (10$^{58}$$d_{8.5}^{2}$ cm$^{-3}$) & 5.4$^{+2.0}_{-2.5}$ & 6.2$^{+1.3}_{-1.7}$ & 5.4$^{+3.4}_{-3.6}$ & 2.7$^{+0.5}_{-0.5}$ & 4.4$^{+1.5}_{-1.1}$ & 1.6$^{+4.4}_{-0.1}$ & 6.0$^{+2.8}_{-1.6}$ & 4.8$^{+9.3}_{-0.7}$ & 2.4$^{+4.0}_{-1.2}$ & 106$^{+139}_{-26}$ \\

$n_{e_s}$ ($f^{-1/2}_s$$d_{8.5}^{-1/2}$ cm$^{-3}$) & 8.4$^{+1.6}_{-1.9}$ & 10.4$^{+1.1}_{-1.4}$ & 8.1$^{+2.6}_{-2.8}$ & 4.8$^{+0.5}_{-0.5}$ & 6.3$^{+1.1}_{-0.8}$ & 10.7$^{+14.7}_{-0.4}$ & 28.6$^{+6.7}_{-3.8}$ & 15.2$^{+14.9}_{-1.1}$ & 4.6$^{+3.8}_{-1.2}$ & 6.6$^{+4.3}_{-0.8}$ \\

$t_{sh_s}$ ($f^{1/2}_s$$d_{8.5}^{1/2}$ kyr) & $>$4.5 & 1.4$^{+1.8}_{-0.6}$ & $>$17.3 & $>$11.8 & $>$2.9 & $>$3.0 & 0.4$^{+0.7}_{-0.1}$ & $>$1.7 & $>$5.0 & $>$35.8 \\

$f_s$  & {\bf 0.21} & 0.43 & 0.55 & 0.52 & 0.30 & 0.47 & 0.56 & 0.38 & 0.46 & 0.47 \\

\cline{1-1}
\textit{Hard component} & \\
\cline{1-1}

 EM$_h$ (10$^{58}$$d_{8.5}^{2}$ cm$^{-3}$) & 1.6$^{+0.4}_{-0.5}$ & 0.7$^{+0.2}_{-0.4}$ & 0.4$^{+0.2}_{-0.3}$ & 0.2$^{+0.1}_{-0.1}$ & 0.8$^{+0.4}_{-0.3}$ & 0.3$^{+0.7}_{-0.1}$ & 0.3$^{+0.2}_{-0.1}$ & 0.4$^{+0.2}_{-0.1}$ & 0.4$^{+0.4}_{-0.2}$ & 11.7$^{+19.0}_{-8.6}$ \\

 $n_{e_h}$ ($f^{-1/2}_h$$d_{8.5}^{-1/2}$ cm$^{-3}$) &  4.5$^{+0.6}_{-0.8}$ & 3.4$^{+0.6}_{-1.1}$ & 2.2$^{+0.4}_{-0.8}$ & 1.4$^{+0.2}_{-0.4}$ & 2.7$^{+0.8}_{-0.4}$ & 4.4$^{+6.1}_{-0.8}$ & 6.3$^{+1.7}_{-1.0}$ & 4.3$^{+1.2}_{-0.4}$ & 1.9$^{+1.0}_{-0.4}$ & 2.2$^{+1.8}_{-0.8}$ \\

$n_{0}$ ($f^{-1/2}_h$$d_{8.5}^{-1/2}$ cm$^{-3}$) & 0.9$^{+0.1}_{-0.2}$ & 0.7$^{+0.1}_{-0.2}$ & 0.4$^{+0.1}_{-0.2}$ & 0.3$^{+0.1}_{-0.1}$ & 0.6$^{+0.2}_{-0.1}$ & 0.9$^{+1.3}_{-0.2}$ & 1.3$^{+0.3}_{-0.2}$ & 0.9$^{+0.3}_{-0.1}$ & 0.4$^{+0.2}_{-0.1}$ & 0.5$^{+0.4}_{-0.2}$ \\

$t_{sh_h}$ ($f^{1/2}_h$$d_{8.5}^{1/2}$ kyr) &  1.9$^{+0.8}_{-0.7}$ & 1.3$^{+0.7}_{-0.6}$ & 1.0$^{+0.5}_{-0.7}$ & 1.2$^{+0.3}_{-0.8}$ & 3.2$^{+1.9}_{-1.2}$ & 2.3$^{+3.5}_{-1.0}$ & 0.5$^{+0.3}_{-0.2}$ & 0.6$^{+0.7}_{-0.2}$ & 2.5$^{+1.5}_{-1.1}$ & 1.2$^{+1.0}_{-0.5}$  \\

$f_h$ & {\bf 0.79} & 0.57 & 0.45 & 0.48 & 0.70 & 0.53 & 0.44 & 0.62 & 0.54 & 0.53  \\

\hline
\end{tabular}
\tablecomments{$EM$: Emission measure, $n_e$: electron density, $n_0$: ambient density, $t_{sh}$: shock age, and $f$: filling factor.}
\end{table*}

It has been previously suggested that the remnant could be either in its early evolutionary stage or in a transition to its Sedov phase (Gotthelf \& Vasisht 1997; Vink \& Kuiper 2006). Under the assumption of a uniform ambient density medium, the swept-up mass is calculated as $M_{sw}$ = ($\frac{4}{3}\pi$$R_s^3$)$\times$1.4$m_p$$n_0$ = 16$^{+13}_{-6}f^{-1/2}_hD_{8.5}^{5/2}$~M$_{\sun}$. For a filling factor $f_h$~=~0.53 (Table~4), $M_{sw}$ $\sim$ 14--40 $D_{8.5}^{5/2}$ M$_{\sun}$. This range implies that the remnant can be in its late free expansion phase or in a transition to the Sedov phase. 

It is also plausible that the SNR may still be in the dense wind in an $r^{-2}$ profile, which would be more realistic for Kes~73 if the explosion originated from a massive star (e.g., SN IIL/b) as proposed in Section 5.4. We discuss below the evolutionary properties of  Kes~73 under both scenarios.

Assuming a free expansion (or Sedov) phase gives a lower (or higher) estimate on the SNR age. For an initial expansion velocity of 5000 km~s$^{-1}$ appropriate for a core-collapse SNR (Reynolds 2008), we infer a free expansion age of $\lesssim$750~$d_{8.5}$ kyr for the remnant. When the swept-up mass ($M_{sw}$) becomes comparable to or exceeds the ejected mass ($M_{ej}$), an SNR enters the adiabatic phase.  If Kes~73 is in its early Sedov phase, we can compute an upper limit to the Sedov age as $t$ = $\eta$$R_s$/$V_s$ using the VSEDOV spectral parameters (Table~3), where $\eta$ depends on the evolutionary stage of the SNR and is equal to 0.4 assuming a uniform blast wave expansion (Sedov 1959). The blast wave velocity $V_s$ is estimated as $V_s$ = (16$k_B$$T_s$/3$\mu$$m_H$)$^{1/2}$, where $\mu$ = 0.604 is the mean mass per free particle for a fully ionized plasma, $k_B$ = 1.38$\times$10$^{-16}$ ergs K$^{-1}$ is Boltzmann's constant and $T_s$ is the post-shock temperature (Sedov 1959).  Assuming full equilibration between the electrons and nuclei (a valid assumption here since our VSEDOV fit gives similar temperatures for the mean temperature of the particles and the electron temperature right behind the shock), we determine the shock velocity using the tabulated temperature from our SEDOV fit  (Table~3) of $kT_h$ = 1.6$^{+0.8}_{-0.7}$~keV. The inferred velocity of the blast wave is $V_s$ = (1.2$\pm$0.3)$\times$10$^3$ km~s$^{-1}$, and the Sedov age is estimated as $t$~=~2.1$\pm$0.5~$d_{8.5}$~kyr for Kes~73. This age is consistent with the previous estimated age of $\leq$2.2 kyr (Gotthelf \& Vasisht 1997; Vink \& Kuiper 2006; Lopez et al. 2011). However, incomplete equilibration at the shock would lead to a higher shock velocity than that estimated above, yielding an even younger age. Studies of electron and ion temperatures in SNR shocks indicate that equilibration is far from complete at such high velocities (Ghavamian et al. 2007), suggesting that Kes~73 is considerably younger than the characteristic age of the associated AXP, $\tau_c \approx 4.7$~kyr, which provides an upper limit for the system age.  To the best of our knowledge, there is no measurement of the braking index which would provide an accurate age estimate for 1E~1841$-$045. 

Based on the Sedov blast wave model in which a supernova with an explosion energy $E$, expands into an ISM of uniform density $n_0$, we can estimate the SNR explosion energy as $E$ = $\frac{1}{2.02}$$R_s^5$$m_n$$n_0$$t_{SNR}^{-2}$ = 3.0$^{+2.8}_{-1.8}$$\times$10$^{50}f_h^{-1/2}d^{5/2}_{8.5}$ ergs, where $m_n$ = 1.4$m_p$ is the mean mass of the nuclei and $m_p$ is the mass of the proton. This value is slightly lower than the canonical value of $10^{51}$~ergs, but is consistent with the explosion energy  of (5 $\pm$ 3)$\times$10$^{50}$~ergs derived by Vink \& Kuiper (2006) assuming a distance of 7.0~kpc.  If full electron-ion equilibration\footnote{As mentioned earlier, the VSEDOV fit to the global SNR spectrum, however, yielded equal electron and ion temperatures for the global SNR spectrum. } has not been achieved in the shock for the Sedov phase, the shock velocity $V_s$ determined from the electron temperature will be a lower limit to the shock temperature $T_s$, which will lead to an underestimation of the SNR's explosion energy. 

Now, we discuss the model treatment of an SNR expanding into the late red-supergiant (RSG) phase wind of its massive progenitor. This is motivated by the progenitor mass estimate inferred from our study (Section 5.4) and follows the expansion of the SNR in an $r^{-2}$ wind profile as discussed in Chevalier (2005) and applied to other SNRs such as G296.1--0.5 (Castro et al. 2011).   In the following discussion however, we will focus on the SNR's expansion in a wind profile in the adiabatic phase, and infer the SNR's expansion velocity and age in comparison to the values estimated from its expansion into a uniform medium (i.e. the Sedov phase described above). The Sedov profiles for an adiabatic blast wave expansion in a power-law density distribution have been solved analytically by Cox and Franco (1981). Chevalier (2005) describes the circumstellar wind density of a SN IIL/b progenitor as $\rho_{CS}$ = $\dot{M}/4\pi r^2 v_w$ $\equiv$ $Dr^{-2}$ and defines the dimensionless quantity $D_{*}$ = $D$/$D_{ch}$, where $D_{ch}$ = 1$\times$10$^{14}$~g~cm$^{-1}$ is the value of the coefficient of the density profile, $\dot{M}$ is the mass loss and $v_w$ is the RSG wind velocity.  The circumstellar mass swept-up by the SNR shock to a radius $R$ is then given by $M_{sw}$ = 9.8$D_*$($R$/5~pc) $M_{\sun}$, with the SNR's blast wave expanding to a radius $R$ = (3$E$/2$\pi D$)$^{1/3}$$t^{2/3}$ during the adiabatic phase of its evolution. The corresponding velocity in this Sedov approximation is then given by $V_s$~=~2$R$/3$t$ (Chevalier 1982; Chevalier 2005; Castro et al. 2011). For Kes~73, we assume here that the hot component arises entirely from the shocked CSM and that the shocked wind fills the full volume of $V$ = 2.9$\times$10$^{58}$~cm$^3$ (which is a reasonable assumption for the adiabatic phase). For an emission measure EM$_h$ = 11.7$^{+19.0}_{-8.6}$$\times$10$^{58}$~$d^{2}_{8.5}$~cm$^{-3}$, we obtain $M_{sw}$ = 62$^{+52}_{-24}$ $f_h^{-1/2}$$d_{8.5}^{5/2}$~M$_{\sun}$.  This value will be subsequently used below to determine the supernova's explosion energy using the equation above for $R$ as a function of $E$, $D$ (or $M_{sw}$), and $t$.

Next, we derive the SNR's age ($t$) and shock speed ($V_s$) in the RSG wind phase. Using the Sedov profiles for an adiabatic blast wave in an $r^{-2}$ wind density profile (Cox \& Franco 1981), the ratio between the average and shock temperature (weighted by $n^2$) in the RSG wind scenario is $\langle T\rangle/T_s$ = 5/7, whereas in the uniform case $\langle T\rangle/T_s$ = 1.276.  Assuming that the X-ray spectrum depends mostly on the average temperature, we approximate the ratio between the shock temperatures and shock speeds in the wind and uniform cases as $T_{s, wind}$/$T_{s, uniform}$ = 1.786 and $V_{s, wind}$/$V_{s, uniform}$ =  $\sqrt{1.786}$, respectively. Now, since $V_s$ = $\eta$ $R$/$t$ where $\eta$=2/5 for the Sedov uniform case and $\eta$=2/3 for the wind adiabatic case, we estimate the ages ratio as $t_{s, wind}$/$t_{s, Sedov}$ = (5/3) /  $\sqrt{1.786}$~=~1.247. Using the above-estimated Sedov velocity and age of (1.2$\pm$10$^3$)$\times$10$^3$~km~s$^{-1}$ and 2.1$\pm$0.5~kyr, respectively, we obtain $V_s$ = (1.6$\pm$0.7)$\times$10$^3$~km~s$^{-1}$ and $t_{wind}$ = 2.6$\pm$0.6~$d_{8.5}$~kyr. By substituting $t_{wind}$ in the equations for $M_{sw}$ as a function of the SNR's age and explosion energy, we can infer  $E$ =  1.1$^{+1.0}_{-0.6}$$\times$10$^{51}$~$f_h^{-1/2}$$d_{8.5}^{5/2}$~ergs, which corresponds to $\sim$1.5$\times$10$^{51}$ ergs for $f_h$=0.53 and a distance of 8.5~kpc.
 
Subsequently, the shock age $t_{sh}$ (the time since the passage of shock) of the two thermal components was calculated using the ionization timescale ($n_et$) values derived from the fits (Table~4), noting that the soft-component is approaching ionization equilibrium in some of the regions (Table~3). We obtained a shock age of $t_{sh, h}$~=~(0.5--3.2) $f_h^{1/2}$$d_{8.5}^{1/2}$ kyr for the hard component, which is more or less consistent with the SNR age.

\subsection{Progenitor mass}
\label{5.4}

Nucleosynthesis studies through X-ray spectroscopy of shock-heated ejecta associated with young SNRs can reveal important information on the nature of their progenitor stars.  It is possible to directly measure the abundances of heavy elements obtained from X-ray spectral fitting. By comparing them to the models of explosive nucleosynthesis yields such as those of Woosley \& Weaver 1995 (hereafter WW95) and Nomoto et al. 2006 (hereafter N06), we can estimate the mass of the progenitor star. Such studies are also valuable for testing and improving the nucleosynthesis models commonly used to fit X-ray spectra. 

We use the abundance ratios relative to Si given by (X/Si)/(X/Si)$_{\sun}$, where X is the ejected mass of any element with respect to Si and with solar values obtained from Anders \& Grevesse (1989). In Figure~6, we show the ratios (with respect to Si) for O, Mg, S, and Fe, obtained from the global SNR spectral fitting, in comparison to the predicted ratios for a range of progenitor masses using the WW95 and N06 models. WW95 specify models A, B, and C for progenitor stars with masses $\ge$30 $M_{\sun}$, where the final kinetic energies at infinity of the ejecta for models B and C are enhanced by a factor of $\sim$1.5 and $\sim$2, respectively, with respect to model A ($KE_{\infty}$ $\sim$ 1.2$\times$10$^{51}$~ergs). The yields from WW95 and N06 are different due to the different treatment of the convective boundary and since mass loss is not included in WW95 models, whereas N06 includes mass loss as a function of metallicity (K. Nomoto 2013, private communication). As a result, the inferences on the progenitor masses are highly dependent on the models used.  When comparing with the WW95 and N06 yields, the abundance ratios are consistent with a 20--30 M$_{\sun}$ progenitor star (Figure~6). As mentioned before, S and Si are overabundant in most of the regions and while the nucleosynthesis models predict a S/Si ratio $\lesssim$1, the data here seem to show a much higher S/Si ratio (although with a large error bar).  A plausible reason could be that, the remnant being young, the reverse shock may not have reached the center yet and hence, the current abundances may not reflect accurately the actual abundances of the heavy elements produced in the supernova explosion. Another factor contributing to the uncertainty in the S abundance is that the S line lies in the energy band where the two components have nearly the same flux. We have also compared our values with lower progenitor masses ($\lesssim$18 M$_{\sun}$), however, the ratios do not match with such lower mass stars. Nevertheless, keeping in mind the error bars and uncertainty in the models, as well as the caveats mentioned above, the existing data and models suggest a progenitor mass $\gtrsim$20 M$_{\sun}$ for Kes~73. 

\begin{figure*}[th]
\includegraphics[width=\textwidth]{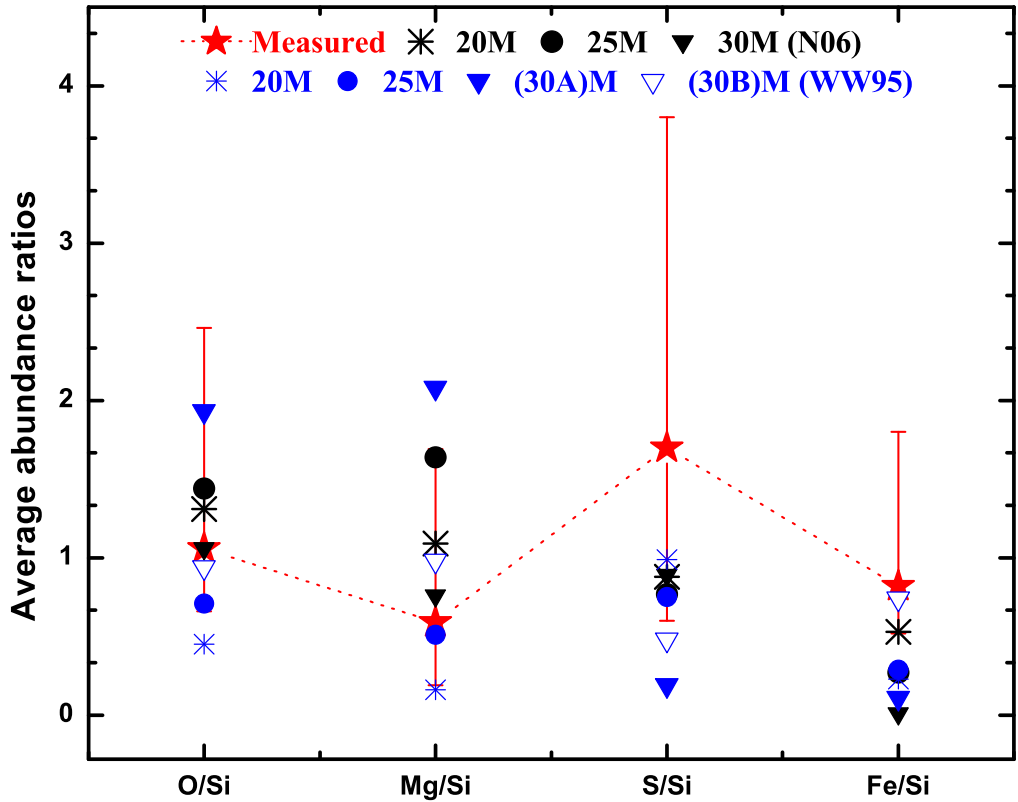}
\caption{Best fit abundances of O, Mg, S, and Fe relative to Si relative to solar. The data points are connected with a dotted line to help guide the eye. Also, over plotted are the predicted relative abundances [X/Si]/[X/Si]$_{\sun}$ from the core-collapse nucleosynthesis models of WW95 and N06. The models 30A and 30B of WW95 differ in their final kinetic energies at infinity of the ejecta, where model B is enhanced by a factor of $\sim$1.5 with respect to model A. See Section 5.4 for details.}
\end{figure*}

Chevalier (2005) suggests a supernova type SN IIL/b for Kes~73, with the remnant running into the RSG wind lost from the progenitor star. We discussed this scenario in Section 5.3. SN IIL/b stars are believed to have progenitor masses in the range of 25--35$M_{\sun}$ (Heger et al. 2003) with X-ray luminosities typically in the range of $\sim$10$^{37}$~ergs~s$^{-1}$. They end their lives as RSGs with extensive mass loss ($\dot{M}$ $\ge$ 3$\times$10$^{-5}$~$M_{\sun}$~yr$^{-1}$ for a wind velocity $v_w$ = 15~km~s$^{-1}$) from the H envelope, leading to the formation of a dense circumstellar region that extends out to 5--7 pc from the progenitor (Chevalier 2005). Interestingly, the radius of Kes~73 ($\sim$6.2~pc) also closely matches the expected size of the SN IIL/b wind bubble, and together with our inferred X-ray luminosity ($\sim$3$\times$10$^{37}$~ergs~s$^{-1}$) and progenitor mass of $\gtrsim$20$M_{\sun}$, the prediction of a SN IIL/b progenitor star for Kes~73 seems reasonable.

Observational studies show that several magnetars are associated with star clusters and by studying their association with these clusters, it is possible to estimate the initial mass of their progenitors. A recent study by Clark et al. (2009), on the discovery of a third massive RSG cluster (RSGC3) located at the base of the Scutum-Crux arm, suggests that the SNR Kes~73 could be physically associated with the RSGC1--3 star formation complex with the location of 1E~1841$-$045 approximately equidistant from the RSG clusters 2 and 3. Therefore, they speculate that the progenitor mass of 1E~1841$-$045 cannot be $\ge$20$ M_{\sun}$ by associating the AXP with the star formation that yielded RSGC1--3. Together with another study performed by Davies et al. (2009) which reported a low mass progenitor star of 17 $\pm$ 2$ M_{\sun}$ for the soft gamma repeater (SGR) 1900+14, these authors suggest that magnetars do not necessarily descend from massive progenitors. However, as mentioned above, based on our X-ray spectroscopic analysis as well as the nucleosynthesis yield models used here for comparison, we estimate a massive progenitor $\ge$20$ M_{\sun}$ for Kes~73. 

Next, we compare the estimated progenitor mass for Kes~73 to that determined for a few other magnetars and high-magnetic field radio pulsars. The progenitor masses of SGR 1806--20 associated with the cluster CI 1806--20, CXOU J164710.2--455216 associated with the cluster Westerlund 1 (Wd 1), and the AXP 1E 1048.1--5937 were estimated as 48$^{+20}_{-8}$ $M_{\sun}$ (Bibby et al. 2008), 40 $\pm$ 5 $M_{\sun}$ (Muno et al. 2006), and 30--40 $M_{\sun}$ (Gaensler et al. 2005), respectively.  The progenitor masses obtained for the two high-B pulsars (J1119--6127 and J1846--0258) also suggests massive progenitors. The X-ray spectroscopic study of SNR G292.2--0.5 associated with the high-B radio and X-ray pulsar J1119--6127 suggests a progenitor mass $\sim$30 $M_{\sun}$ (Kumar et al. 2012). The SNR Kes 75, associated with the high-magnetic field X-ray pulsar J1846--0258 (which recently showed magnetar-like bursts and spectral softening; Gavriil et al. 2008; Kumar \& Safi-Harb 2008), was  suggested to have a Wolf-Rayet progenitor (Morton et al. 2007).  In summary, the above mentioned studies imply a range of progenitor masses for the highly magnetized neutron stars, with the majority implying very massive progenitors. However, any conclusion on the factors deciding the origin and evolutionary properties of these sources requires further studies in X-rays and other wavelengths. In particular, in the X-rays, future high-resolution spectroscopic studies are needed to obtain accurate abundance measurements and to perform plasma diagnostics of the individual lines to confirm the temperature of the ejecta as well as that inferred for the interstellar/circumstellar environment in which Kes~73 is expanding.

\section{Summary and Conclusions} 
\label{6}

In this paper, we have performed the first detailed imaging and spectral analysis of the SNR Kes~73, using all archival \textit{Chandra} and \textit{XMM-Newton} data, to determine the intrinsic properties of the supernova explosion and the physical properties of the remnant.  The main results are summarized as follows: 

\begin{enumerate}
\item
The high resolution X-ray images confirm a spherical morphology of $\sim$5$\arcmin$ size with several clumpy and knotty structures, and bright diffuse emission originating from the SNR interior. The multi-wavelength morphology of the remnant in the radio, infrared, and X-rays shows the western limb to be much brighter than all other regions. The infrared image spatially correlates well with the X-ray image while the radio shell appears to extend slightly beyond the X-ray emission and displays a thin filamentary feature along the edges, which likely represents the location of the forward shock. 
\item
The spectra obtained from different diffuse emission regions are best fit with a VPSHOCK+VPSHOCK model. The western limb of the SNR showed a slightly higher column density given by $N_H$ = 3.3$^{+0.1}_{-0.3}$$\times$10$^{22}$~cm$^{-2}$. The soft component is dominated by enhanced metal abundances mainly for Si and S in all the regions, with plasma temperatures of 0.3--0.5~keV and ionization timescales $>$10$^{12}$~cm$^{-3}$~s (except for regions~2 and 7). The hard component is dominated by solar abundances, with plasma temperatures of 1.1--1.7~keV and ionization timescales of (0.5--3.2)$\times$10$^{11}$~cm$^{-3}$~s. This indicates that the soft component plasma for most regions has reached ionization equilibrium earlier than the plasma associated with the hard component. 
\item
The global X-ray emission from  Kes~73 is best described by a two-component VSEDOV+VPSHOCK model, with a column density $N_H$ = 2.6$^{+0.4}_{-0.3}$$\times$10$^{22}$~cm$^{-2}$ and the soft and hard components characterized by plasma temperatures of 0.5$^{+0.1}_{-0.2}$ keV ($n_et_s$ $>$ 6.1$\times$10$^{12}$~cm$^{-3}$~s) and 1.6$^{+0.8}_{-0.7}$ keV ($n_et_h$ = 0.8$^{+0.2}_{-0.1}$ $\times$ 10$^{11}$~cm$^{-3}$~s), respectively. The presence of enhanced abundances in the soft component suggests that this component is dominated by shock-heated ejecta, while the hard component characterized mostly by solar abundances is dominated by shocked interstellar/circumstellar material. We have also refined the SNR age ranging between 750~$d_{8.5}$ yr for the free expansion phase (assuming an expansion velocity of 5000 km~s$^{-1}$) and 2100~$d_{8.5}$~yr assuming a Sedov phase of evolution.  The Sedov phase yields a shock velocity of (1.2 $\pm$ 0.3)$\times$10$^{3}$~km~s$^{-1}$, an explosion energy of $E$ = 3.0$^{+2.8}_{-1.8}$$\times$10$^{50}f_h^{-1/2}d^{5/2}_{8.5}$~ergs, and a swept-up mass of 16$^{+13}_{-6}f_h^{-1/2}D_{8.5}^{5/2}$~M$_{\sun}$ under the assumption of an explosion in a uniform ambient medium. Considering Kes~73 to be still expanding into the dense wind of its late phase RSG evolution, we infer a shock velocity of (1.6$\pm$0.7)$\times$10$^3$~km~s$^{-1}$, an age of 2.6$\pm$0.6~$d_{8.5}$~kyr, and explosion energy $E$ = (1.1$^{+1.0}_{-0.6})$$\times$10$^{51}$~$f_h^{-1/2}$$d_{8.5}^{5/2}$~ergs. These derived values are consistent with the predictions of a very massive progenitor for Kes~73.  
\item
Although the abundances are not well-constrained, the abundance ratios, when compared to core-collapse nucleosynthesis models, suggest a progenitor mass $\gtrsim$20$~M_{\sun}$ for Kes~73. A much deeper exposure with existing X-ray missions, and high-resolution spectroscopy with the Soft X-ray Spectrometer of \textit{ASTRO-H} (Takahashi et al. 2012), are needed for an accurate measurement of the abundances and shock velocities, and to further motivate the development of the existing nucleosynthesis models.
\end{enumerate}

\acknowledgments
We thank Ken$^{\prime}$ichi Nomoto for helpful discussions on the nucleosynthesis yield models for core-collapse supernovae and the referee for useful comments that helped improved the paper. This research made use of NASA's Astrophysics Data System (ADS) and of NASA's HEASARC maintained at the Goddard Space Flight Center (GSFC). S. Safi-Harb acknowledges support by a Discovery Grant from the Natural Sciences and Engineering Research Council of Canada (NSERC), the Canada Research Chairs (CRC) program, the Canada Foundation for Innovation, the Canadian Institute for Theoretical Astrophysics, and the Canadian Space Agency. POS acknowledges support from NASA Contract NAS8-03060.

\textit{Facilities: CXO (ACIS), XMM-Newton (EPIC)}

\end{document}